\documentclass[aps,reprint,superscriptaddress,preprintnumbers]{revtex4-2}
\usepackage{amsmath}
\usepackage{xr-hyper}
\usepackage{hyperref}
\usepackage{isotope}
\usepackage{mhchem}
\usepackage{xcolor}
\usepackage{graphicx}
\usepackage{mathrsfs}
\usepackage{dcolumn}
\usepackage{bm}
\usepackage{cases}
\usepackage{multirow,booktabs}
\usepackage{makecell}
\usepackage{amsmath}
\usepackage{amssymb}
\usepackage{sidecap}
\usepackage{graphicx}
\usepackage{here}
\usepackage{braket}
\usepackage{adjustbox}

\usepackage{tikz}
\usepackage{float}
\usepackage{tabularx}
\usepackage{soul}
\usepackage[normalem]{ulem}
\hypersetup{
  breaklinks=true,
  colorlinks=true,
  linkcolor=blue,
  urlcolor=blue,
  citecolor=blue
}

\usepackage{xcolor}
\definecolor{pastelgray}{rgb}{0.81, 0.81, 0.77}
\definecolor{beaublue}{rgb}{0.9, 0.9, 0.93}

\usepackage{todonotes}

\begin{document}

\title{A physics-embedded Bayesian neural network for predicting the energy dependence of fission product yields with fine structures}

\author{Jingde Chen}
\author{Yuta Mukobara}%
\author{Kazuki Fujio}%
\altaffiliation[Current affiliation: ]{Los Alamos National Laboratory, USA}
\affiliation{%
 Department of Transdisciplinary Science and Engineering, School of Environment and Society, Institute of Science Tokyo, 2-12-1 Ookayama, Meguro, 152-8550, Tokyo, Japan
}%


\author{Satoshi Chiba}
\affiliation{%
Professor Emeritus, Institute of Science Tokyo, 2-12-1 Ookayama, Meguro, 152-8550, Tokyo, Japan
}%

\author{Tatsuya Katabuchi}
\author{Chikako Ishizuka}
  \email{ishizuka.c.686a@m.isct.ac.jp}
\affiliation{
 Laboratory for Zero-Carbon Energy, Institute of Integrated Research, Institute of Science Tokyo, Meguro, 152-8550, Tokyo, Japan
}%


\date{\today}

\begin{abstract}
We present a physics-embedded Bayesian neural network (PE-BNN) framework that integrates fission product yields (FPYs) with prior nuclear physics knowledge to predict energy-dependent FPY data with fine structure. 
\textcolor{black}{By incorporating a phenomenological shell-related input feature with an excitation-energy damping term, the PE-BNN captures both fine structures and global energy trends.}
The combination of this physics-informed input with hyperparameter optimization via the Watanabe-Akaike Information Criterion (WAIC) significantly enhances predictive performance. 
\textcolor{black}{Our results show that the PE-BNN framework is well suited to observables whose systematic features can be encoded as model inputs, and that it remains consistent with known shell-related trends and independently measured prompt-neutron systematics.}
%
%
\end{abstract}

\maketitle

\section{Introduction}
Nearly 90 years after the discovery of nuclear fission, understanding its reaction mechanisms remains a major challenge in nuclear physics~\cite{Abdurrahman+2024}. At the same time, fission plays a vital role in both fundamental research and applications, including nuclear reactors, reactor neutrinos~\cite{Andriamirado+2025}, superheavy element synthesis, and astrophysical nucleosynthesis~\cite{Wanajo+2024}. Among key observables, fission yields provide crucial insights into the fission process. Neutron-induced fission yields, in particular, have been extensively measured and compiled into standard nuclear data libraries \cite{ENDF,jendl23,JEFF}, making them indispensable for theoretical and engineering studies.

\textcolor{black}{A well-documented case is neutron-induced fission of $^{235}$U, where decades of experiments have led to a precise understanding of its fission yield structure, minimizing discrepancies between evaluated nuclear data libraries. However, neither recent nuclear-physics models nor empirical nuclear-data models can fully reproduce fine structure variations across isotopes and incident neutron energies. Since 2018, machine learning techniques have been applied to fission yield predictions, \textcolor{black}{successfully capturing global trends \cite{Chen01,global_1,BNN-FPY,global_3,global_4}}. 
%
However, a systematic and predictive description of fine structures \cite{fine_structure} remains an open challenge.}

\textcolor{black}{Another key issue is the energy dependence of the fission yields. Current nuclear data libraries provide yields at only three specific neutron energies, 2.53$\times 10^{-8}$ MeV, 0.5 MeV, and 14 MeV, chosen for engineering needs. Intermediate values are traditionally estimated by linear interpolation, but recent experiments have revealed significant deviations \cite{Pierson_235U_238U_232Th,NAIK2015_238U,NAIK2013_238U,NAIK_2016_232Th,Bhatia_235U_238U_239Pu,Gooden_235U_238U_239Pu}, underscoring the need for more accurate predictive methods \cite{burnup-monitor,99mTc}. Limited access to mono-energetic neutron sources further restricts comprehensive experimental investigations, making robust theoretical models essential.}

\textcolor{black}{
To address these challenges, we developed a Bayesian neural network (BNN) model \cite{neal96} integrated with advanced nuclear physics models. Our previous study \cite{Chen01} demonstrated that data augmentation through weighted training improves the ability of BNNs to capture fine structure details. Here, we refine this approach by introducing a hybrid modeling strategy that incorporates multiple theoretical calculations, including our dynamical models, to improve predictions of global yield structures. \textcolor{black}{All data used for training in the present work were selected from sources currently regarded as reliable, regardless of whether they are experimental, evaluated, or theoretical. While existing physics-based models are highly successful in some respects, their predictive performance is not always sufficient for all fission-yield observables; therefore, we utilize only well-established physical information and adopt a machine-learning framework to improve predictions where current models remain limited.} Additionally, we introduce a novel shell factor inspired by nuclear mass prediction \cite{NIU201848} to improve interpretability and precision.
}
\textcolor{black}{The fine structures appearing in fission product yields after prompt neutron emission reflect not only shell effects but also $\beta$-decay contributions. To avoid the complexity of explicitly treating $\beta$-decay, evaluated, experimental, and theoretical yields after prompt neutron emission are adopted as the training data.} \textcolor{black}{In the present work, the PE-BNN is trained to predict post-neutron independent mass yields. Some of the experimental datasets used for validation, however, correspond to cumulative yields measured by $\gamma$-ray spectroscopy; accordingly, the comparisons in this paper are intended to assess global energy-dependent trends and mass-dependent structures rather than strict point-by-point equality.}

\textcolor{black}{
Our results reveal that the shell factor is crucial for reproducing fine structure variations. \textcolor{black}{Moreover, our predicted energy dependence aligns well with independent experimental data and shows strong consistency with prompt neutron emission measurements \cite{mueller1981numerical}}. We further confirm the robustness and generality of our findings through validation in multiple isotopes and cross-validation techniques \cite{Waic1}}.
%
\textcolor{black}{These findings address two central issues in fission-yield studies: the reproduction of fine structures and the description of energy dependence within a unified framework. The present model bridges nuclear-physics modeling and statistical learning and provides a practically useful framework for applications in reactor physics, nuclear forensics, and astrophysical nucleosynthesis.}


\textcolor{black}{\section{Computational Methods}\textcolor{black}{To predict the Fission Product Yields (FPYs), we have developed Bayesian neural networks (BNNs)~\cite{Chen01}.}
In the BNNs, the posterior over model parameters $\mathcal{W}$ given data $\mathcal{D} = \{(\boldsymbol{x}_i, \boldsymbol{y}_i)\}_{i=1}^n$ is computed using Bayes' theorem:
\begin{equation}
    p(\mathcal{W}|\mathcal{D}) = \frac{p(\mathcal{D}|\mathcal{W})p(\mathcal{W})}{p(\mathcal{D})} = \frac{p(\mathcal{D}|\mathcal{W})p(\mathcal{W})}{\int p(\mathcal{D}|\mathcal{W})p(\mathcal{W})d\mathcal{W}}.
\end{equation}
Here, $p(\mathcal{W})$ is the prior (often zero-mean Gaussian), and $p(\mathcal{D}|\mathcal{W})$ is the likelihood derived from the data. The denominator $p(\mathcal{D})$ known as evidence, is typically intractable.
In the BNN model, learning is performed via Maximum a Posteriori estimation, which seeks the parameters $\hat{\mathcal{W}}$ that maximize the posterior:
\begin{align}
    \hat{\mathcal{W}} &= \arg\max_{\mathcal{W}} \log p(\mathcal{W}|\mathcal{D}) \\
    &= \arg\max_{\mathcal{W}} \left( \log p(\mathcal{D}|\mathcal{W}) + \log p(\mathcal{W}) + \text{const} \right).
\end{align}
\textcolor{black}{The No-U-Turn Sampler (NUTS) \cite{hmc-nuts} 
is used to seek $\hat{\mathcal{W}}$ from posterior distribution. 
\textcolor{black}{NUTS automatically adjusts the step size, preventing inefficient random walks and improving computational efficiency in Hamiltonian Monte Carlo (HMC) \cite{HMC}.}} We perform the NUTS computations using the NumPyro \textcolor{black}{library} \cite{numpyro}.} After training the BNN, predictions for new data $\boldsymbol{X}'$ are made by marginalizing over the posterior \textcolor{black}{\cite{equation5}}. 
\textcolor{black}{Here, $Y'$ is predicted mass yield}: 
\begin{equation}
    p(\boldsymbol{Y}^{'}|\boldsymbol{X}^{'},\mathcal{D}) = \int _{\mathcal{W}} p(\boldsymbol{Y}^{'}|\boldsymbol{X}^{'}, \mathcal{W})p(\mathcal{W}|\mathcal{D})d\mathcal{W}. \label{eq:prediction}
\end{equation}

\begin{figure}[b]
\centering
\includegraphics[width=1.0\hsize]{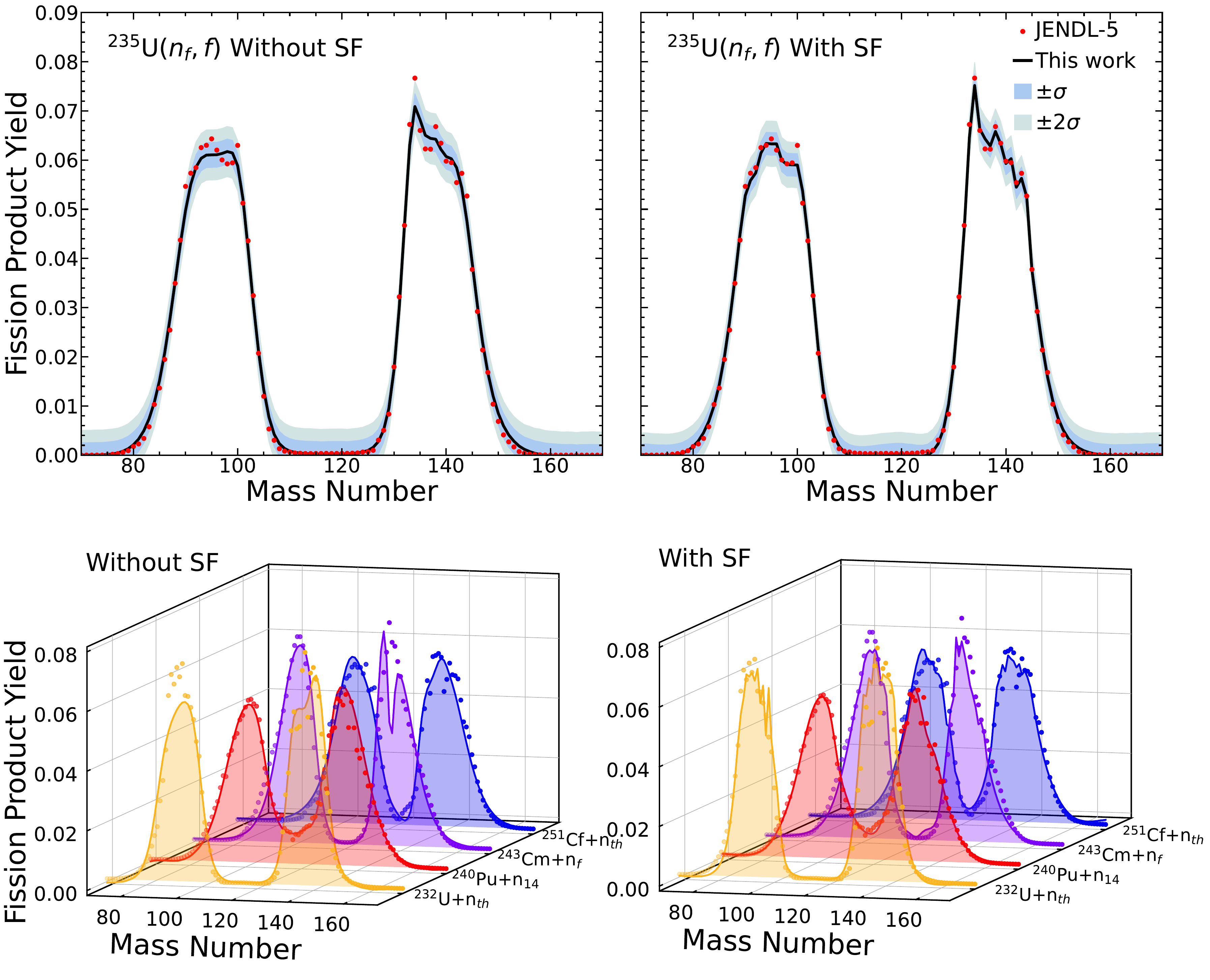} 
\caption{
\textcolor{black}{Comparison of reproduced results (upper) and validated results of the FPYs (lower) with and without the shell factor. The neutrons n$_{th}$, n$_f$ , and n$_{14}$ refer to thermal, fast and 14 MeV incident neutron energies, respectively.
}}
\label{fig:235U_SF_prediction}
\end{figure}

\textcolor{black}{Our BNN model adopts a two-hidden-layer architecture.} In our previous work~\cite{Chen01}, we set the input as $\boldsymbol{x} = (A, Z_n, A_n, E)$, \textcolor{black}{where $A$ is the mass number of the fission products, $Z_n$ and $A_n$ are the charge and mass numbers of the compound nucleus, \textcolor{black}{and} $E$ is the excitation energy, defined as the sum of the neutron binding energy and the incident neutron energy}. To incorporate more physics, we define a new input term, \textcolor{black}{\textit{shell factor}} $\mathcal{SF}$ to reflect shell effects near scission, given by:
\begin{align}
    \mathcal{SF} &= \exp\left(-\frac{E}{kT}\right) \nonumber \\ 
    &\times \left(W_{1} (\mathcal{N}_{1}+\mathcal{N}_{4})+W_{2}(\mathcal{N}_{2}+\mathcal{N}_{5}) 
    +W_{3}(\mathcal{N}_{3}+\mathcal{N}_{6})\right), 
\end{align}
where the $E$ is the \textcolor{black}{ excitation} energy, $kT$, $W_{1}$, $W_{2}$ and $W_{3}$ are adjustable hyperparameters.
\textcolor{black}{
Considering the influence of the shell effects, including double shell closure and deformed shell structures, we assign the parameters as follows: $W_1$ = 4, $W_2$ = 2, and $W_3$ = 1.} The $\exp\left(-\frac{E}{kT}\right)$ represents 
\textcolor{black}{
the Boltzmann factor where $k$ and $T$ denote the Boltzmann constant and temperature}, 
implying that higher incident neutron energies lead to a reduction in shell effects. 
\textcolor{black}{The terms $\mathcal{N}_{i}(A)$ for $i=1,2,3$, follow Gaussian distributions centered on $A=134$, $140$, and $144$, where $A=134$ is associated with the double shell closure, and $A=140$ and $144$ correspond to deformed shell structures, highlighting a preference for fine structure in the heavy fragment region. The corresponding terms $\mathcal{N}_{4}(A;\mu=A_{n}-134)$, $\mathcal{N}_{5}(A;\mu=A_{n}-140)$, and $\mathcal{N}_{6}(A;\mu=A_{n}-144)$ mirror these features on the light fragment side.} \textcolor{black}{Additionally, the standard deviation $\sigma$ is set to 1 for all Gaussians $\mathcal{N}_{1},\mathcal{N}_{2},…,\mathcal{N}_{6}$.}
%
\textcolor{black}{The effective damping scale is set to $kT = 1.5$~MeV, guided by the Fermi-gas relation $kT = \sqrt{E_{\text{int}}/a}$, with $a$ representing the level density parameter and $E_{\text{int}}$ the internal energy. This assignment is further substantiated by multidimensional Langevin dynamical simulations, which predict a characteristic scission temperature of roughly 1.5~MeV for actinide low-energy fission~\cite{Ishizuka_2017}.}

\textcolor{black}{The training set comprises 80\% JENDL-5~\cite{jendl23} (6554 data points), experimental FPY data from EXFOR~\cite{Pierson_235U_238U_232Th,NAIK2015_238U,NAIK_2014_232Th,NAIK2013_238U,NAIK_2016_232Th,Bhatia_235U_238U_239Pu,Gooden_235U_238U_239Pu} (989 data points), and theoretical models~\cite{fujio2023talys,fujio2024langevin} (2062 data points)}, combined with prioritized weighting~\cite{Chen01}. 
\textcolor{black}{
As those training data inherently include multi-chance fission effects, the predictions also incorporate these contributions implicitly.} \textcolor{black}{It should be noted that this study focuses on the evolution of mass yields as a function of incident neutron energy. While the experimental data from EXFOR primarily consist of cumulative fission yields (CFY) and a limited set of independent fission yields (IFY), the quantitative discrepancies between CFY/IFY and the total mass yield are marginal. These differences are significantly smaller than the intrinsic prediction errors of the machine learning model. Furthermore, any data points where the deviation between CFY/IFY and the mass yield exceeded a reasonable threshold,
i.e., points differing from the library value by more than 2$\sigma$, 
were manually filtered during the preprocessing stage.}
The remaining 20\% of JENDL-5 is used for validation, as the most reliable data. \textcolor{black}{Although the training dataset is predominantly composed of independent fission yields, and the PE-BNN model natively outputs predictions for these independent yields, the energy-dependent experimental data available for energy-dependent validation consist primarily of cumulative fission yields. Consequently, our benchmark comparisons are evaluated against these experimental cumulative fission yields. Furthermore, because the training data do not incorporate experimental uncertainties, the predictive uncertainty bounds of the PE-BNN—when compared against experimental cumulative yields that report measurement errors—are not propagated from empirical data, but are instead intrinsically derived from the Bayesian inference process.}

Model selection is based on the Watanabe–Akaike information criterion (WAIC)~\cite{Waic1},
\begin{align}
    \text{WAIC} &= T_{N} + \frac{V_{N}}{N}, \\
    T_{N} &= -\frac{1}{N} \sum_{i=1}^{N} \log \mathbb{E}_{\mathcal{W}}[p(X_{i}|\mathcal{W})], \\
    V_{N} &= \sum_{i=1}^{N} \left( \mathbb{E}_{\mathcal{W}}[\log p(X_{i}|\mathcal{W})^2] - \mathbb{E}_{\mathcal{W}}[\log p(X_{i}|\mathcal{W})]^2 \right),
\end{align}
where $\mathbb{E}_{\mathcal{W}}$ denotes expectation over the posterior of parameters $\mathcal{W}$. \textcolor{black}{$T_{N}$ represents empirical loss, while \textcolor{black}{$V_{N}$} is the functional variance, \textcolor{black}{which helps penalize the use of more parameters in BNNs with larger variance in their predictions.}} \textcolor{black}{Since WAIC is asymptotically equivalent to the k-fold cross-validation~\cite{k-CV} at 
$N \rightarrow \infty$~\cite{watanabe2}, it can be computed from a single model fit, significantly reducing computational cost for model selection.} 
\textcolor{black}{Then we optimized the BNN by selecting a double hidden-layer configuration with 11-11 neurons.}
The selected BNN is then used to predict FPY data, benchmarked against experimental results~\cite{Chapman_235U_238U,Glendenin_235U,Selby,Laurec_233U_235U_239Pu}.

\section{Results and Discussion}
%
\textcolor{black}{
As demonstrated in the following results, the proposed model successfully reproduces and predicts both the global and fine structures of FPY, a task that has long challenged conventional approaches. An analysis of the contribution of each newly incorporated input revealed that the inclusion of theoretical calculation results was primarily responsible for improvements in the global structure, whereas the shell factor exerted a distinct and exclusive influence on the fine structure.}
%



\textcolor{black}{Figure~\ref{fig:235U_SF_prediction} shows the reproduction and validation results with and without the shell factor (SF). Including the SF improves the reproduction of fine structures in both the heavy and light peaks for $^{235}$U. Validation for $^{232}$Th + $n_{th}$, $^{243}$Cm + $n_f$, and $^{251}$Cf + $n_{th}$ reveals similar fine structures, while $^{240}$Pu + $n_{14}$ does not, regardless of SF inclusion. This prediction aligns with our physical intuition, \textcolor{black}{as fine structures tend to disappear with increasing incident neutron energy.} In all cases, with and without SF, both models capture the global trend: as the mass number of the fissioning nucleus and the incident neutron energy increase, the heavy peak in asymmetric fission remains largely unchanged, while the light peak shifts toward the heavy peak.}

\textcolor{black}{
It is also worth clarifying the methodological distinction between the present PE-BNN approach and widely used semi-empirical models such as GEF. 
In GEF, fine structures and their energy dependence are effectively incorporated through a large set of adjustable parameters optimized to reproduce available experimental data. 
In contrast, the PE-BNN framework encodes phenomenological shell information explicitly as an input feature and constrains the learning process through statistically grounded model selection. 
As a result, the present approach provides a transparent and reproducible description of how fine structures evolve with incident neutron energy, without relying on iterative parameter retuning for individual reaction systems.
}

\begin{figure}[b]
\centering
\includegraphics[width=1.0\hsize]{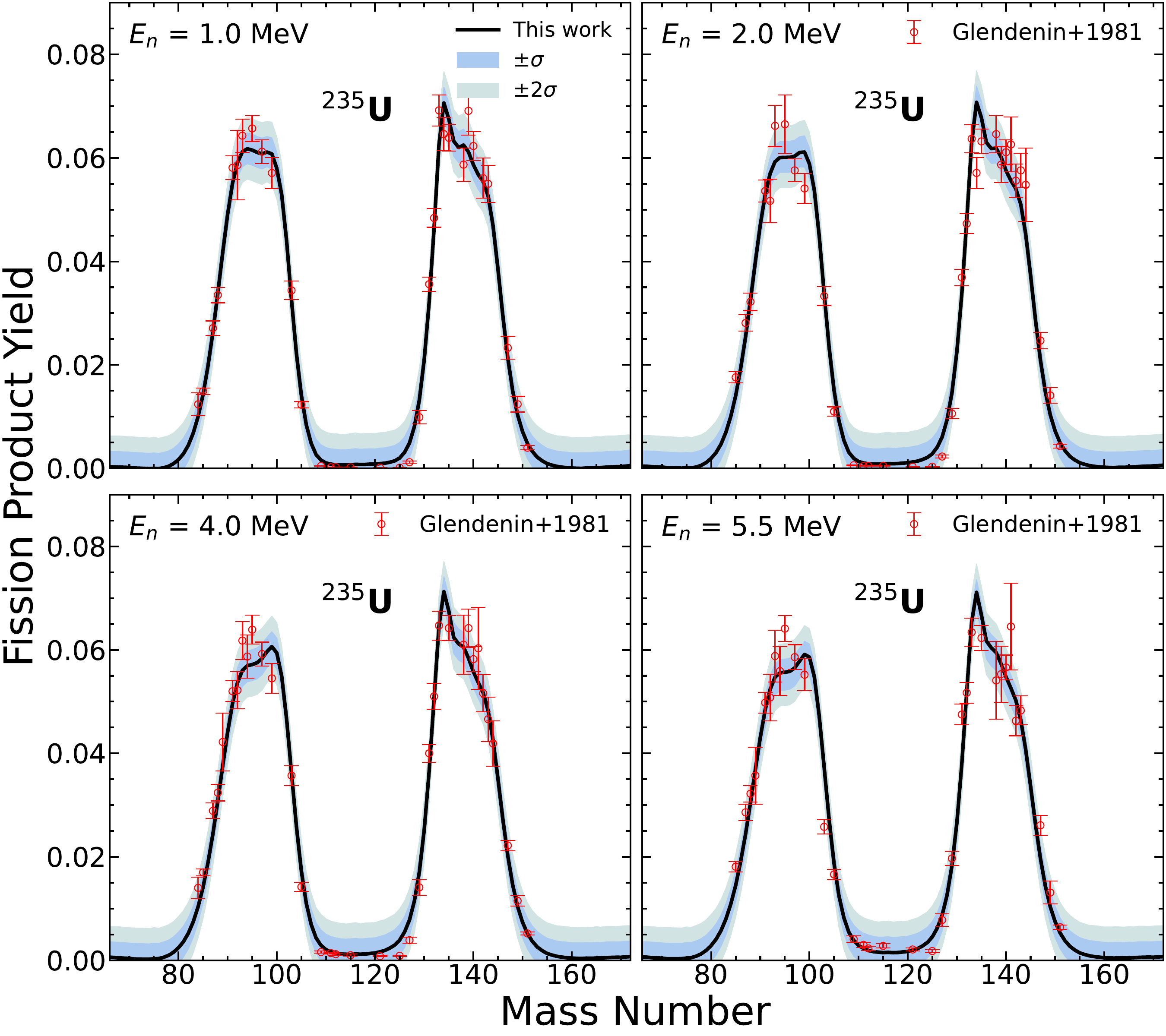}
\caption{Comparison between the \textcolor{black}{FPY(A) after prompt neutron emission but before $\beta$-decay predicted} by the BNN model and \textcolor{black}{the experimental CFY data \cite{Glendenin_235U}.}}
\label{fig:with_without_shell}
\end{figure}

\begin{figure}
\centering
\includegraphics[width=1.0\hsize]{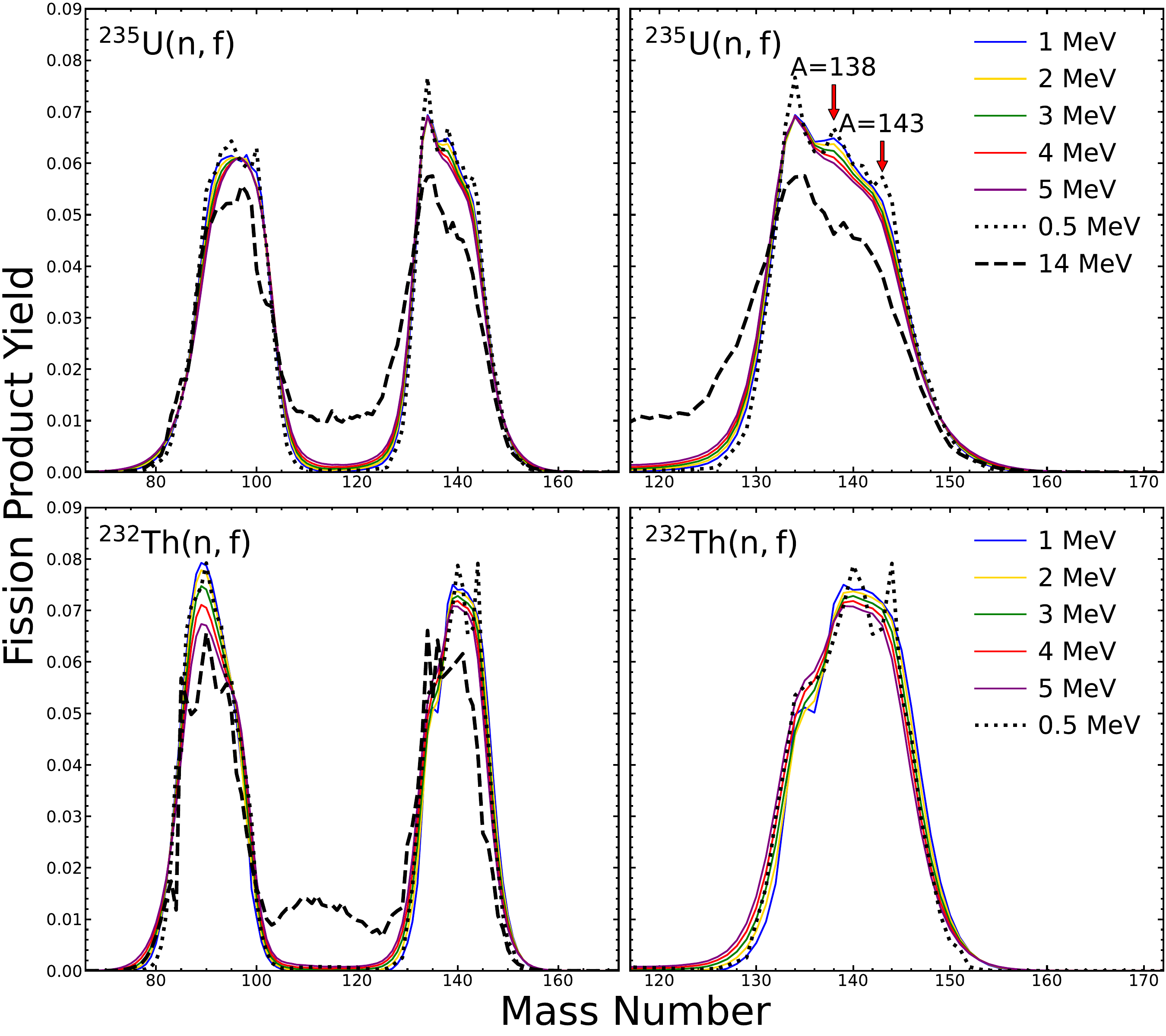}
\caption{\textcolor{black}{The FPY(A)s of $^{235}$U and $^{232}$Th 
mass yield obtained by summing the JENDL-5 independent yields over $Z$}
(0.5 MeV with dotted line and 14 MeV with dashed line) and BNN FPY(A) predictions (colored solid lines)
in the case of our previous model~\cite{Chen01} only with data augmentation without SF.  
In the lower right \textcolor{black}{$^{232}$Th} data, 
\textcolor{black}{which are also obtained by summing the 14 MeV independent-yield from JENDL-5,} 
is omitted to plot due to its large uncertainty.
\textcolor{black}{In the right panels, independent mass yields for the heavy peak are shown.}}
\label{fig:explaining_1A}
\end{figure}
\begin{figure}[t]
\centering
\includegraphics[width=1.0\hsize]{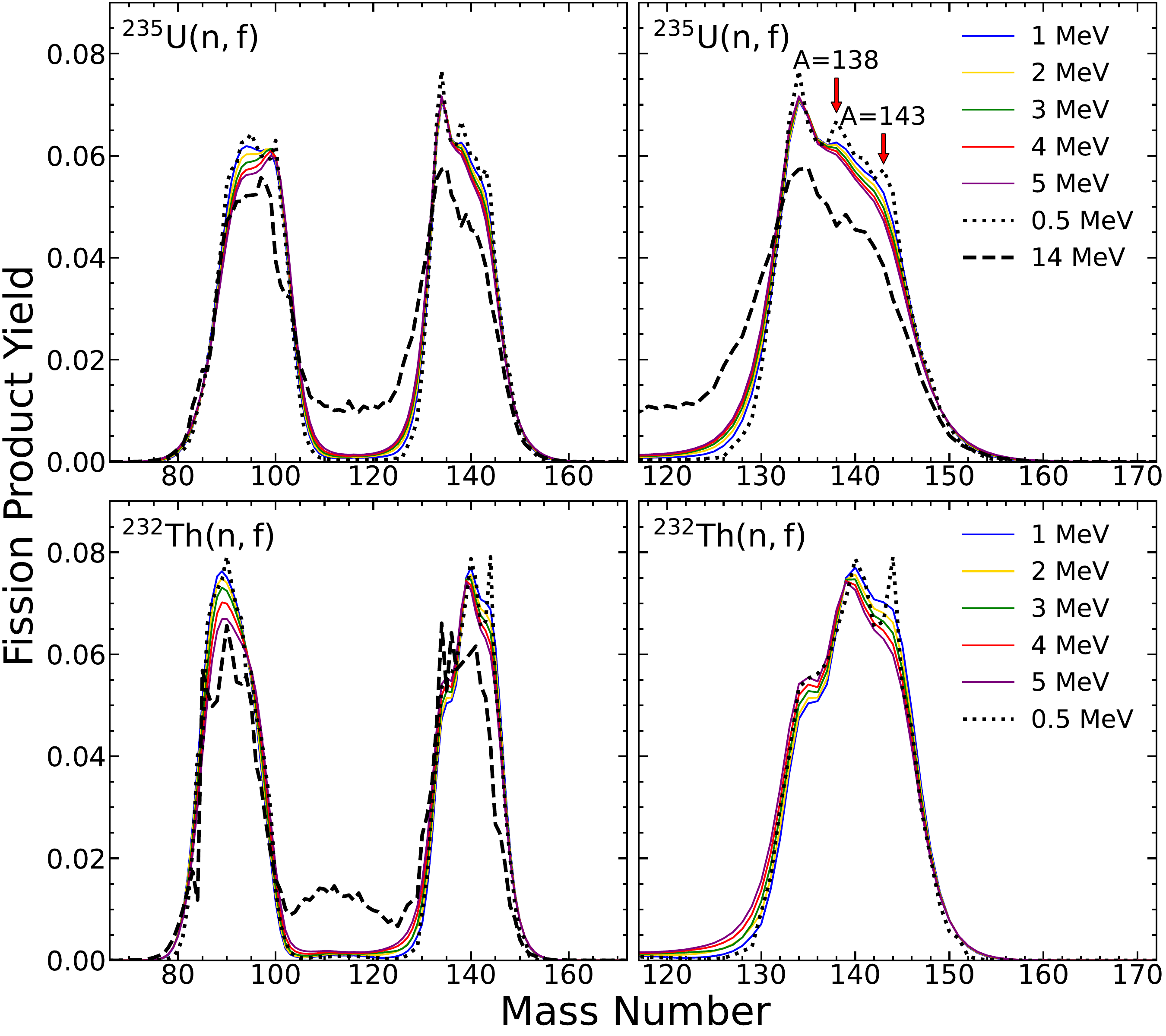}
\caption{The FPYs of $^{235}$U and $^{232}$Th from JENDL-5 (0.5 MeV and 14 MeV) and BNN predictions (solid lines) 
\textcolor{black}{in the case of present model with SF}. 
In the lower right \textcolor{black}{$^{232}$Th} data, the 14 MeV yield from JENDL-5 is omitted to plot due to its large uncertainty. \textcolor{black}{In the right panels, independent mass yields for the heavy peak are shown.}}
\label{fig:explaining_1B}
\end{figure}
\textcolor{black}{To evaluate the impact of the shell factor, we calculated the log-likelihood ($\log \mathcal{L}$) for the training and testing datasets in Fig.~\ref{fig:235U_SF_prediction}:
\begin{align}
    \log \mathcal{L} = -\frac{N \log(2 \pi \sigma^2)}{2} - \frac{\sum_{i=1}^{N} (y_i - f(\boldsymbol{x_{i}};\mathcal{W}))^2}{2 \sigma^2},
\end{align}
where $N$ is the number of data points, $y_i$ the true value, and $f(\boldsymbol{x_i};\mathcal{W})$ the BNN output. \textcolor{black}{In the optimal BNN model with a double hidden-layer configuration of 11-11 neurons, the shell factor boosted log-likelihood by 21.3\% (train) and 34.6\% (test).} Across architectures from 5-5 to 20-20, it improved training and testing performance by 19.6\% and 29.4\% on average.}

%
\textcolor{black}{
We examine our prediction accuracy in neutron-induced fission reactions of $^{235}$U. Figure~\ref{fig:with_without_shell} illustrates a comparison between the BNN predictions and experimental FPY data~\cite{Glendenin_235U} at each energy. The results indicate that the BNN predictions reside within the $\pm 1\sigma$ and $\pm 2\sigma$ uncertainty bands \textcolor{black}{(corresponding to the 68.3\% and 95.5\% credible intervals)}, \textcolor{black}{demonstrating systematical agreement with all experimental values at incident neutron energies $E_{n}$ = 1.0 - 5.5 MeV,} including the fine structures present in the two global peaks.}

%
%
\textcolor{black}{In Figs. \ref{fig:explaining_1A} and \ref{fig:explaining_1B}, we also display the behaviors of the predicted FPYs for $E_{n}$ = 1.0 - 5.0 MeV between two evaluated datasets from JENDL-5
to focus on the influence of the shell effect.}
\textcolor{black}{The present analysis focuses on the low-to intermediate-energy region where shell effects are known to play a significant role. At higher incident neutron energies, where multi-chance fission dominates and shell effects are strongly damped, a more explicit treatment of individual fission channels would be required.}
\textcolor{black}{
In those energy regions, the energy-dependent fission yields below 5 MeV are of particular importance for practical applications such as reactor criticality evaluations \cite{multichance_fission}, so that we carefully examined the behavior of our predicted results.}
\textcolor{black}{Both} figures illustrate a well-established trend: as the neutron incident energy increases, the asymmetric fission product yield decreases, while the symmetric fission product yield increases. Those predicted behaviors are consistent with the global tendency suggested by the classic Brosa's fission model~\cite{brosa}. 
%
%

Furthermore, the right panels of Figs.~\ref{fig:explaining_1A} and ~\ref{fig:explaining_1B} emphasize how the fine structure in the heavy asymmetric mass region varies with \textcolor{black}{incident energies}. 
\textcolor{black}{
For $^{235}$U, the energy dependence of the attenuation of the peaks at $A=138$ and $A=143$ shows a clear sensitivity to the inclusion of the shell factor (SF). The prediction results indicate that, in the heavy asymmetric component, the peaks at $A=138$ and $A=143$ rapidly disappear with increasing incident neutron energy, while the peak at $A=134$ persists. This behavior suggests a hierarchy in the underlying shell stabilization effects. Specifically, the peak at $A=134$ is strongly stabilized by the influence of the doubly magic configuration around $A=132$ ($Z=50$, $N=82$), making it particularly robust against increasing excitation energy. In contrast, the peaks at $A=138$ and $A=143$, which are associated with deformed shell closures, are more strongly damped and vanish as the incident neutron energy increases.}

\textcolor{black}{
Without the SF, the shell effect associated with $A=143$ remains dominant, indicating that this structure is weakly damped with increasing incident energy. When the SF is included, however, the relative importance shifts toward the shell structure around $A=138$, suggesting a different hierarchy of shell effects governing the energy evolution of the heavy-fragment yields.
In the light-fragment region, calculations without the SF exhibit an almost Gaussian-like behavior, with shell effects mainly localized around $A \approx 100$. Once the SF is introduced, energy-dependent fine structures emerge over a broader mass range, indicating that shell-driven correlations are transferred to the light-fragment side through the coupled evolution of the fission system. As a consequence, the energy dependence near $A \approx 100$ becomes qualitatively different from that of neighboring mass numbers, reflecting a nonuniform damping of shell effects with increasing excitation energy.}

\textcolor{black}{
Similarly, the prediction results for $^{232}$Th reveal an intriguing trend.
The influence of the SF is more pronounced than in $^{235}$U; without the SF, shell effects are largely suppressed, resulting in smooth mass-yield distributions, while including the SF generates clear fine structures in both the heavy- and light-fragment regions.
Importantly, the stabilization of the $A=134$ peak and the energy-dependent attenuation of the peaks at $A=138$ and $A=143$ emerge naturally from the trained network, without explicitly imposing such behaviors during training. This indicates that the PE-BNN captures physically meaningful correlations embedded in the training data, rather than relying on direct interpolation of individual mass yields. These results suggest that the present framework provides a physically interpretable description of shell-effect damping with increasing excitation energy, bridging data-driven learning and established nuclear-structure systematics.
}

\begin{figure}
\centering
\includegraphics[width=0.9\hsize]{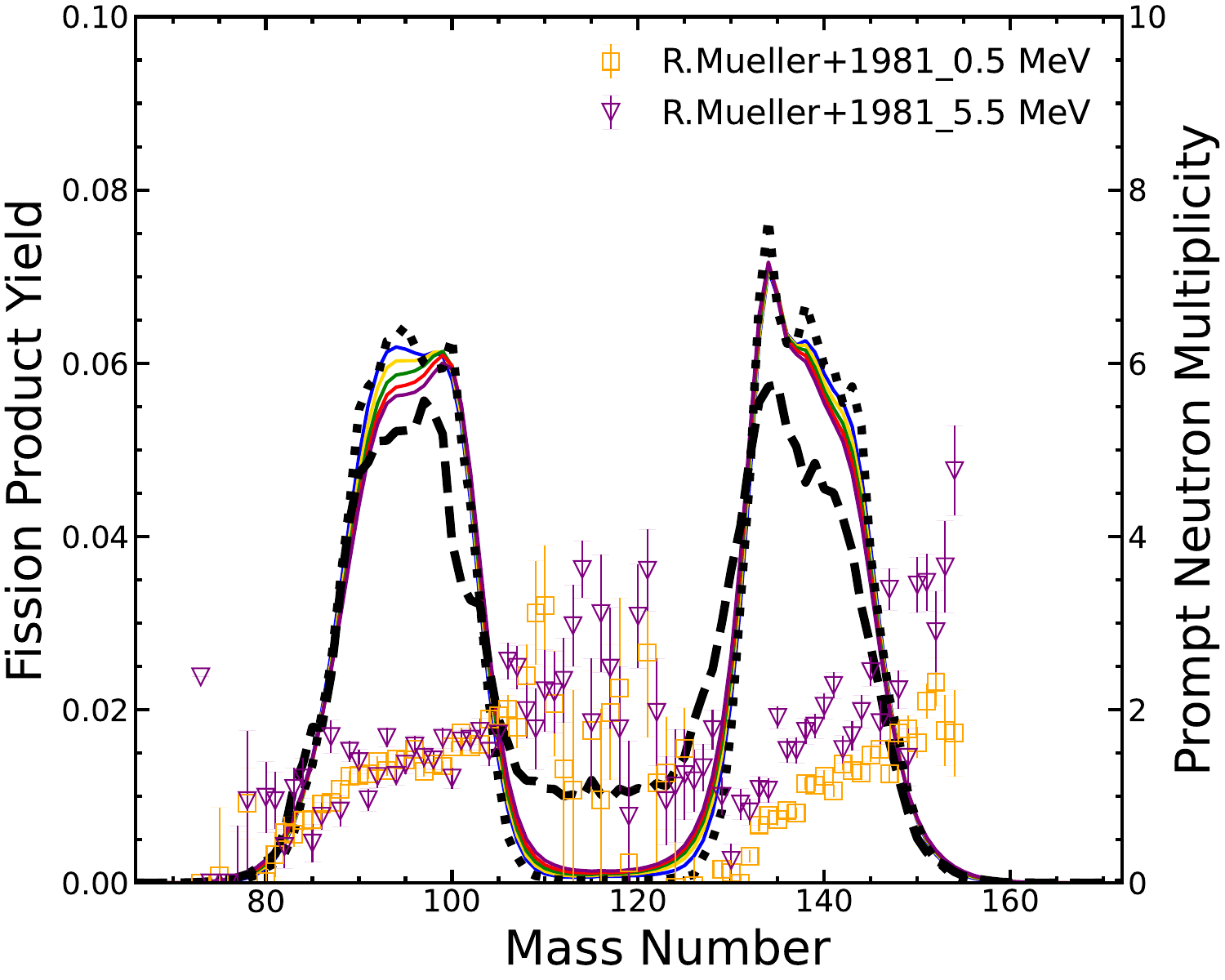}
\caption{The energy dependence of fission product yields and prompt neutron multiplicity for $^{235}$U \textcolor{black}{in the case of
selective learning of nuclides with high practical demand}. Colors of solid lines are the same as in \textcolor{black}{Fig.}~\ref{fig:explaining_1B}.}
\label{fig:explaining_2}
\end{figure}

\textcolor{black}{
Figure~\ref{fig:explaining_2} presents a direct comparison between the energy dependence of fission product yields (FPYs) and the corresponding prompt neutron multiplicities for $^{235}$U(n,f) reactions. The figure provides an important physical reference for interpreting the mass-dependent trends observed in the FPY predictions.
A clear contrast is observed between light and heavy fission fragments. In the light-fragment region, the prompt neutron multiplicity remains nearly constant between $E_n = 0.5$ and 5.5~MeV, indicating that the number of emitted neutrons does not change significantly with incident neutron energy. Consistent with this behavior, the corresponding FPY structures exhibit weak energy dependence, and their ridge positions remain nearly fixed.
In contrast, the heavy-fragment region shows a pronounced increase in prompt neutron multiplicity with increasing incident neutron energy. This increase implies stronger neutron emission at higher excitation energies, resulting in a reduction of post-neutron fragment masses. As a consequence, the FPY ridge structures shift toward lower mass numbers as $E_n$ increases. The correlation between increasing neutron multiplicity and the systematic displacement of FPY ridges suggests a direct physical connection between neutron emission and the observed energy-dependent evolution of the mass-yield distributions.}

%

%

\begin{figure*}[th]
\centering
\includegraphics[width=1.00\hsize]{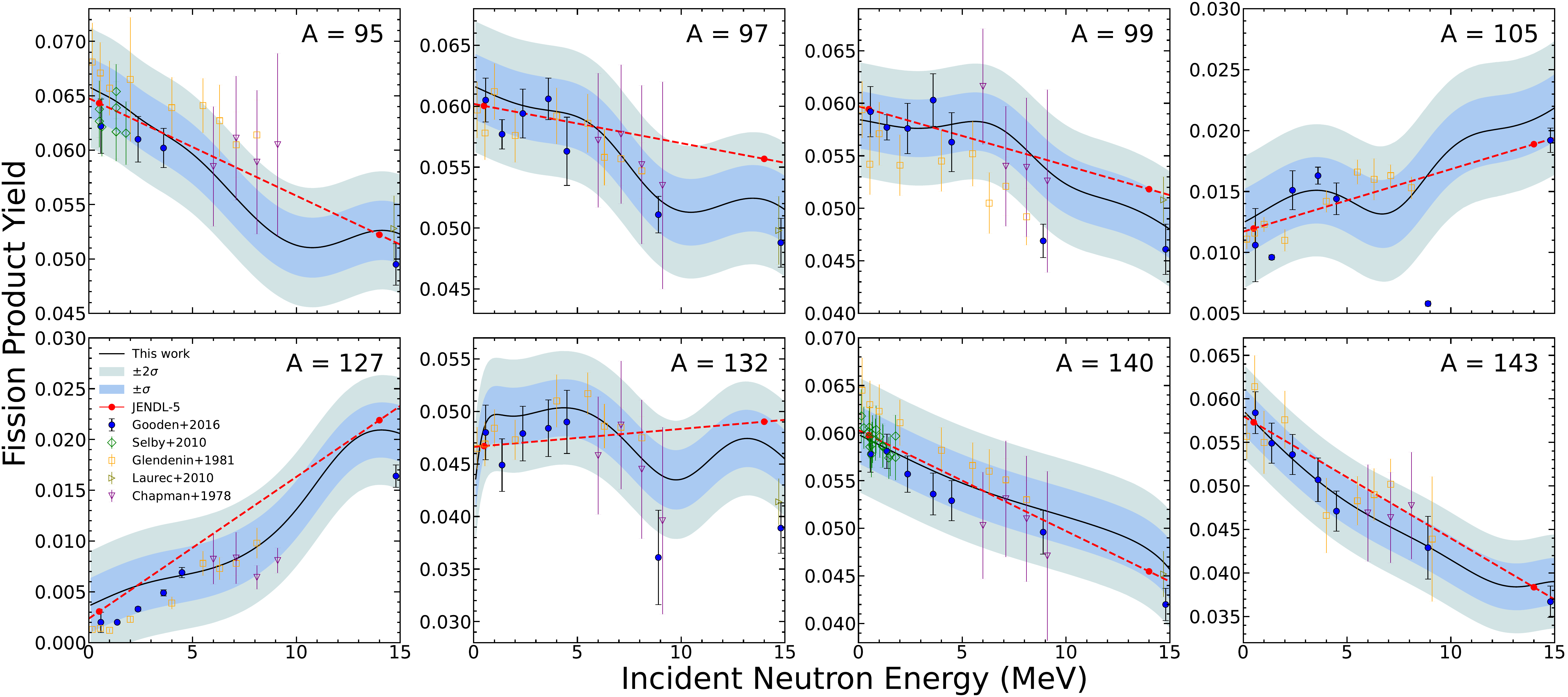}
\caption{\
Black solid lines in each panel,
\textcolor{black}{which are predicted mass yields after prompt neutron emission but before $\beta$-decay,} 
illustrate the energy dependence of \textcolor{black}{independent mass} yields \textcolor{black}{for the neutron-induced fission of $^{235}$U (from $A = 95$ to $143$), within the $\pm 1\sigma$ and $\pm 2\sigma$ uncertainty bands.}
Experimental cumulative yields~\cite{Glendenin_235U,Selby, Laurec_233U_235U_239Pu,Chapman_235U_238U} are shown for validation, \textcolor{black}{where $A = 95$ to $143$ correspond to the specific isotopes $^{95}$Zr, $^{97}$Zr, $^{99}$Mo, $^{105}$Ru, $^{127}$Sb, $^{132}$Te, $^{140}$Ba, and $^{143}$Ce respectively.} The data \textcolor{black}{represented by filled circles}~\cite{Gooden_235U_238U_239Pu} were utilized for model training. The red dashed line denotes a linear interpolation of the JENDL-5 evaluation between 0.5 MeV and 14 MeV.}
\label{fig:Energy_dependence}
\end{figure*}

\textcolor{black}{
Figure~\ref{fig:Energy_dependence} presents a mass-resolved comparison between the PE-BNN predictions and experimental data for selected fission-product yields in the light ($A=95$--105) and heavy ($A=127$--143) asymmetric mass regions \textcolor{black}{for the neutron-induced fission of $^{235}$U}. The horizontal axis represents the incident neutron energy $E_n$ (MeV), and the vertical axis gives the fission-product yield for each selected mass chain. The black solid lines denote the PE-BNN predictions, and the shaded bands indicate the associated Bayesian credible intervals. The red dashed lines represent linear interpolation of JENDL-5 between 0.5 and 14 MeV. Filled symbols indicate data used for training, whereas open symbols represent validation data that were not included during training.}

\textcolor{black}{
For the selected mass chains shown in Fig.~6, the difference between cumulative and independent yields is expected to be small in the present incident-energy range. This is because the dominant $\beta$-decay feeding contributions are limited within the relevant timescales, so that the cumulative yields closely follow the underlying independent \textcolor{black}{mass yields}. Therefore, the comparison between experimental cumulative yields and the PE-BNN predictions of independent \textcolor{black}{mass} yields is justified for discussing global energy-dependent trends and mass-dependent structures.}


\textcolor{black}{
It should first be emphasized that prompt-neutron observables were not used in the training of the PE-BNN model. The training dataset consisted exclusively of FPY information, without including any prompt-neutron multiplicities or neutron-emission systematics. Therefore, the consistency discussed below between the predicted FPY behavior and known prompt-neutron trends should be understood as an emergent manifestation of physical correlations embedded in the FPY data themselves, rather than as a consequence of imposed constraints.
}

\textcolor{black}{
Figure~\ref{fig:Energy_dependence} summarizes the energy dependence of FPYs in both the light ($A=95$--105) and heavy ($A=127$--143) asymmetric mass regions and provides a unified physical interpretation of the behaviors observed in Figs.~\ref{fig:explaining_1B} and~\ref{fig:explaining_2}. In the light-fragment region, the peak positions remain nearly unchanged as the incident neutron energy increases, while the overall distributions gradually broaden. This behavior is consistent with the weak energy dependence observed for nuclides around $A \approx 99$--105 in Fig.~\ref{fig:explaining_2}, where the FPYs exhibit nearly flat slopes. In particular, below $E_n \approx 5$~MeV, isotope-dependent trends are clearly observed: $A=95$ shows a steep \textcolor{black}{decrease}, $A=97$ exhibits a more moderate slope, and $A=99$ remains nearly unchanged. This tendency suggests that isotopes approaching the doubly magic configuration $N=Z=50$ are less sensitive to incident neutron energy. The isotope at $A=105$ also shows a gradual increase with energy. These features are naturally interpreted as a consequence of the nearly energy-independent prompt-neutron multiplicity in light fragments, which preserves the post-neutron peak locations even as the distribution widths increase.
}

\textcolor{black}{
In contrast, the heavy-fragment region exhibits a systematic shift of the ridge structure toward lower mass numbers with increasing incident neutron energy, as clearly seen in Fig.~\ref{fig:Energy_dependence}. This behavior is consistent with the stronger energy dependence observed for heavier nuclides in Fig.~\ref{fig:explaining_2}. Starting from $A=127$, the FPYs generally increase with incident energy, whereas the doubly magic nucleus at $A=132$ maintains a nearly flat slope. The slope at $A=140$ is steeper than that at $A=143$, indicating stronger shell effects around $A=140$, consistent with recent findings in the actinide region \cite{nature1,explanation_huo}. The ridge displacement toward lower masses can be physically understood as a consequence of increasing prompt-neutron multiplicity in heavy fragments: enhanced neutron emission at higher excitation energies reduces post-neutron fragment masses, resulting in a gradual shift of the FPY ridges toward smaller $A$ values.
}

\textcolor{black}{
These contrasting behaviors between light and heavy fragments provide a unified explanation for the energy-dependent broadening of the asymmetric peaks observed in Fig.~\ref{fig:explaining_1B}. With increasing incident neutron energy, the widths of both fragment groups increase, while systematic peak-position shifts appear only in the heavy-fragment region.
}
\textcolor{black}{
Taken together, the consistent behaviors observed in Figs.~\ref{fig:explaining_1B}--\ref{fig:Energy_dependence} indicate that physically meaningful correlations between mass-yield evolution and neutron-emission behavior emerge directly from FPY-only training. This demonstrates that the PE-BNN framework goes beyond simple statistical interpolation and provides a physically interpretable description of energy-dependent fission dynamics.
}

\begin{figure*}[th]
\centering
\includegraphics[width=1.00\hsize]{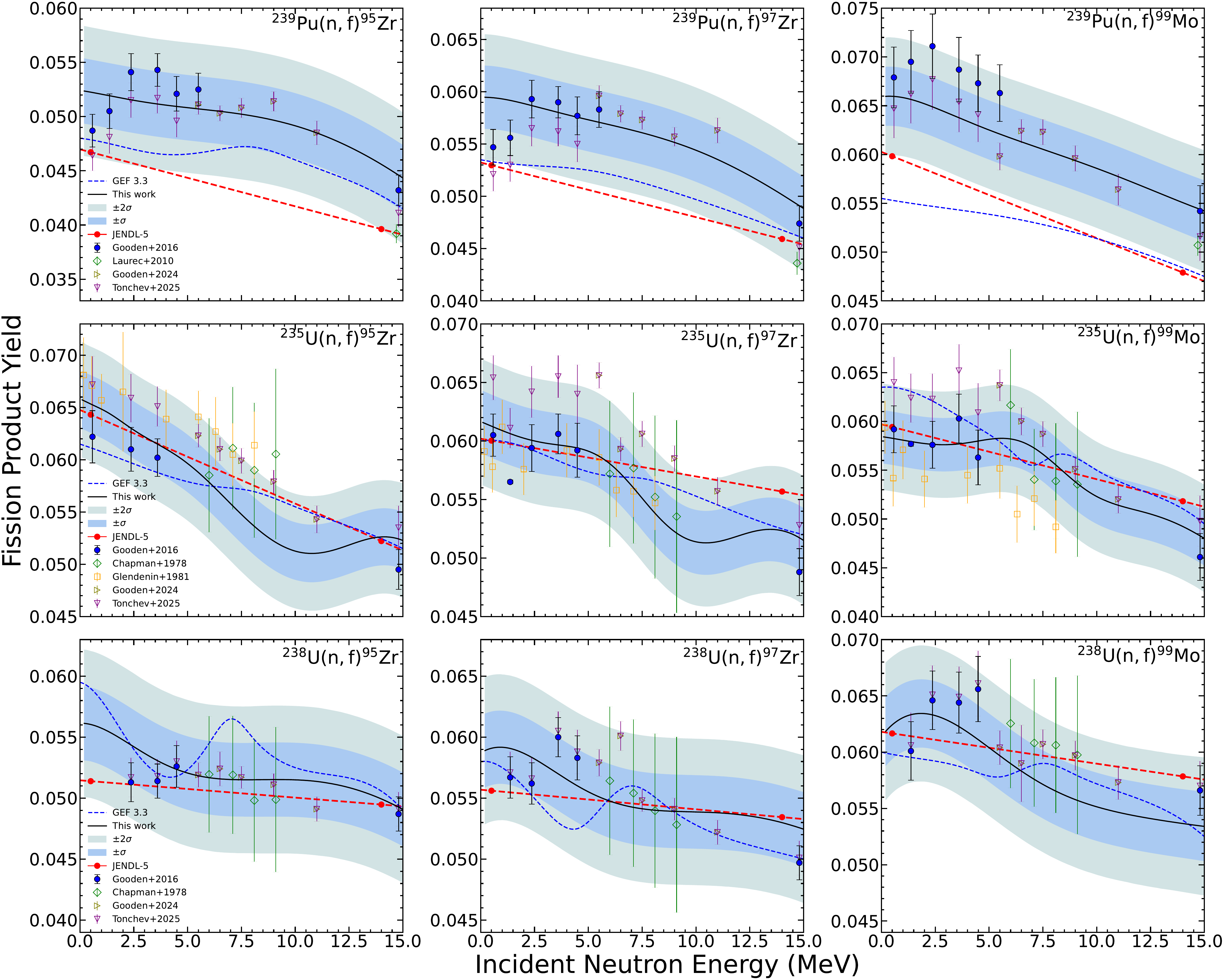}
\caption{\
Black solid lines in each panel, 
\textcolor{black}{which are predicted mass yields after prompt neutron emission but before $\beta$-decay,} 
display the energy dependence of fission mass yields from $A = 95$ to $A = 99$ within the $\pm 1\sigma$ and $\pm 2\sigma$ uncertainty bands.
Experimental cumulative yields~\cite{Glendenin_235U,Selby, Laurec_233U_235U_239Pu,Chapman_235U_238U} are shown for validation, while data 
\textcolor{black}{with filled circles} from~\cite{Gooden_235U_238U_239Pu} were used for training. The red dashed line shows a linear interpolation of JENDL-5 between 0.5 MeV and 14 MeV. \textcolor{black}{The blue dash-dotted line is the GEF 3.3 taken from the reference~\cite{Tonchev2025}.}}
\label{fig:FMY_Predictions_light}
\end{figure*}

\begin{figure*}[th]
\centering
\includegraphics[width=1.00\hsize]{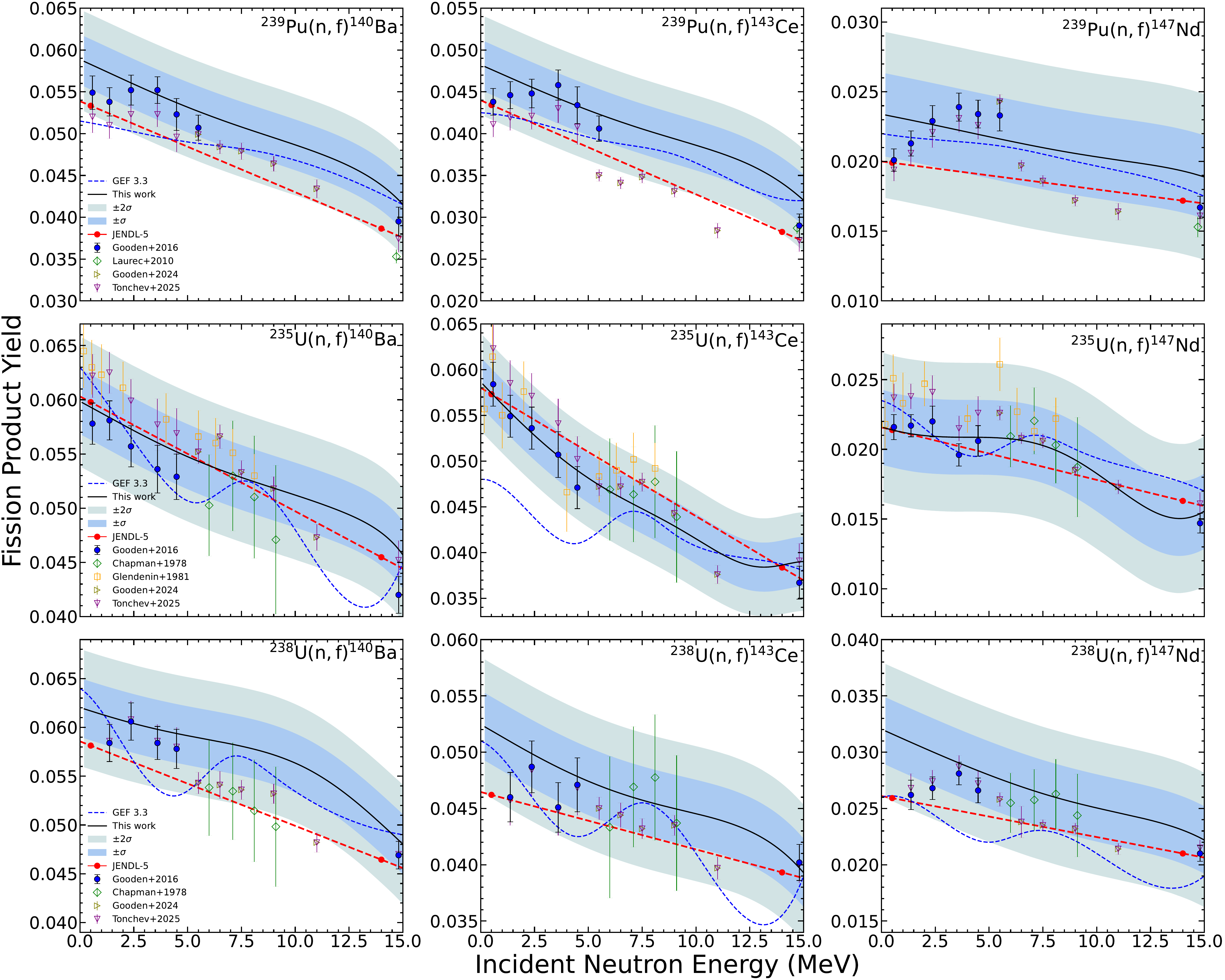} \caption{\textcolor{black}{Similar to Fig.~\ref{fig:FMY_Predictions_light}.}
Black solid lines in each panel,
\textcolor{black}{which are predicted mass yields after prompt neutron emission but before $\beta$-decay,} 
display the energy dependence of fission mass yields from $A = 140$ to $A = 147$within the $\pm 1\sigma$ and $\pm 2\sigma$ uncertainty bands. 
\textcolor{black}{The experimental data are CFY.}}
\label{fig:FMY_Predictions_heavy}
\end{figure*}


\textcolor{black}{
Figures~\ref{fig:FMY_Predictions_light} and~\ref{fig:FMY_Predictions_heavy} present a comparison between the PE-BNN predictions and experimental data for energy-dependent fission-product yields. 
The horizontal axis represents the incident neutron energy $E_n$ (MeV), while the vertical axis shows the fission product yield as a function of fragment mass number $A$. 
The solid black curves correspond to the PE-BNN predictions of independent mass yields, and the shaded bands indicate the associated Bayesian credible intervals representing the model uncertainty.
Dashed or dotted curves denote reference model calculations based on GEF version 2023.3.3 (GEF-3.3), whose values were extracted from the graphical data reported in Tonchev \textit{et al.}~\cite{Tonchev2025}.}

\textcolor{black}{
Experimental data points are shown using symbols with error bars. 
Filled symbols indicate data used for training the PE-BNN model, whereas open symbols represent validation data that were not included during training. 
It should be noted that the experimental values correspond to cumulative fission-product yields measured via $\gamma$-ray spectroscopy, while the PE-BNN predictions represent independent mass yields. 
Therefore, the comparison emphasizes global energy-dependent trends rather than point-by-point equality.}

\textcolor{black}{
The cumulative-yield data used for comparison were taken from the \textcolor{black}{recent} high-precision measurements of Tonchev \textit{et al.}~\cite{Tonchev2025}, which provide systematic energy-dependent FPY data over $E_n = 0.5$--14.8~MeV. 
Importantly, these datasets were not used in training the PE-BNN model. \textcolor{black}{They were instead reserved as an out-of-sample benchmark to test whether the framework can reproduce the observed energy dependence without retraining on the newly available dataset.} 
The good agreement observed in Figs.~\ref{fig:FMY_Predictions_light} and~\ref{fig:FMY_Predictions_heavy} therefore supports the predictive capability of the present framework and indicates that the model captures underlying physical correlations governing the energy evolution of fission yields.} 

\textcolor{black}{
Figures~\ref{fig:FMY_Predictions_light} and~\ref{fig:FMY_Predictions_heavy} provide a broader validation context and a physics-consistent interpretation enabled by the PE-BNN framework. Although the model was trained only on selected FPY datasets and without explicit constraints from prompt-neutron observables or cumulative-yield systematics, it reproduces the global energy evolution observed in independent experimental data. 
The emergence of consistent ridge shifts, peak broadening, and mass-dependent energy sensitivities indicates that the network captures physically meaningful structures embedded in the data rather than performing simple interpolation. 
These results suggest that the PE-BNN framework provides a data-driven pathway toward revealing latent correlations in fission dynamics and offers a physically interpretable bridge between statistical learning and underlying reaction mechanisms. }


\section{Conclusion}
A physics-embedded Bayesian neural network (PE-BNN) framework was developed that synergistically integrates FPYs with prior nuclear physics knowledge to reproduce and predict the energy dependence of FPY data with fine structures. By incorporating a phenomenological shell-related input feature with excitation-energy damping, the PE-BNN reproduces fine structures and the energy-dependent trends of FPY in a physically interpretable manner. \textcolor{black}{Notably, the predicted results, including their energy dependence, are consistent with shell-related trends and with independently reported prompt neutron multiplicities.} 
\textcolor{black}{
This consistency is particularly significant because prompt neutron multiplicities were not included in the training data, indicating that physically meaningful energy-dependent fission systematics can emerge directly from FPY-only learning.}
Our findings demonstrate that the combination of two key strategies---(1) introducing a single input that encapsulates the phenomenological characteristics governing the underlying nuclear fission processes, and (2) optimizing hyperparameters via WAIC\textcolor{black}{---}can significantly enhance the predictive performance of the BNN model while simultaneously enforcing physically meaningful constraints on its behavior.
%
\textcolor{black}{The PE-BNN framework provides a systematic way to incorporate physically motivated input features into Bayesian neural networks for fission yield analysis. 
By combining phenomenological shell information with statistically grounded model selection, the present approach offers a practically useful framework for improving energy-dependent fission product yields in nuclear data evaluations and reactor physics applications.}

\textcolor{black}{
Further comparison with independent experimental datasets, including cumulative-yield data over a broad incident-neutron-energy range, also supports the predictive capability of the present framework beyond simple interpolation between evaluated energy points. These results reinforce the view that the PE-BNN approach provides not only improved predictive performance but also a physically interpretable description of the underlying fission dynamics.} \textcolor{black}{Looking ahead, a key direction for future development concerns the consistent treatment of independent and cumulative fission yields within the training dataset. In the present work, the training set comprises a mixture of evaluated independent fission yields from JENDL-5, experimental independent fission yields, and experimental cumulative fission yields, an amalgamation that, while practically effective, introduces a degree of conceptual inconsistency. To address this limitation, two complementary strategies are envisaged. The first is to extend the PE-BNN architecture to simultaneously predict both independent and cumulative fission yields as distinct output quantities, for instance by augmenting the input feature space with a yield-type identifier and explicitly encoding the physical relationship between IFY and CFY—namely, the radioactive decay chains connecting precursor and product nuclides—as an additional physics-informed constraint within the network. The second, and potentially more fundamental, approach is to develop a robust unfolding procedure that converts experimentally measured cumulative fission yields into independent fission yields prior to training, thereby constructing a fully homogeneous IFY-based training dataset. Such an unfolding would require reliable decay-chain information and careful uncertainty propagation, but would eliminate the yield-type mixing at its source. Either strategy would place the energy-dependent predictions on a more rigorous footing and further strengthen the physical interpretability of the framework.}

\section*{Acknowledgments}
\textcolor{black}{J. Chen and C. Ishizuka acknowledge support by 
the MEXT Innovative Nuclear Research and
Development Program Grant Number JPMXD0222682620, entitled
``Fission product yields predicted by machine learning technique at unmeasured energies and its influence on reactor physics assessment" entrusted to the Tokyo Institute of Technology. J. Chen and Y. Mukobara thank JST SPRING, Japan Grant Number JPMJSP2180.}

\nocite{*}

\bibliography{apssamp}

\begin{thebibliography}{42}%
\makeatletter
\providecommand \@ifxundefined [1]{%
 \@ifx{#1\undefined}
}%
\providecommand \@ifnum [1]{%
 \ifnum #1\expandafter \@firstoftwo
 \else \expandafter \@secondoftwo
 \fi
}%
\providecommand \@ifx [1]{%
 \ifx #1\expandafter \@firstoftwo
 \else \expandafter \@secondoftwo
 \fi
}%
\providecommand \natexlab [1]{#1}%
\providecommand \enquote  [1]{``#1''}%
\providecommand \bibnamefont  [1]{#1}%
\providecommand \bibfnamefont [1]{#1}%
\providecommand \citenamefont [1]{#1}%
\providecommand \href@noop [0]{\@secondoftwo}%
\providecommand \href [0]{\begingroup \@sanitize@url \@href}%
\providecommand \@href[1]{\@@startlink{#1}\@@href}%
\providecommand \@@href[1]{\endgroup#1\@@endlink}%
\providecommand \@sanitize@url [0]{\catcode `\\12\catcode `\$12\catcode `\&12\catcode `\#12\catcode `\^12\catcode `\_12\catcode `\%12\relax}%
\providecommand \@@startlink[1]{}%
\providecommand \@@endlink[0]{}%
\providecommand \url  [0]{\begingroup\@sanitize@url \@url }%
\providecommand \@url [1]{\endgroup\@href {#1}{\urlprefix }}%
\providecommand \urlprefix  [0]{URL }%
\providecommand \Eprint [0]{\href }%
\providecommand \doibase [0]{https://doi.org/}%
\providecommand \selectlanguage [0]{\@gobble}%
\providecommand \bibinfo  [0]{\@secondoftwo}%
\providecommand \bibfield  [0]{\@secondoftwo}%
\providecommand \translation [1]{[#1]}%
\providecommand \BibitemOpen [0]{}%
\providecommand \bibitemStop [0]{}%
\providecommand \bibitemNoStop [0]{.\EOS\space}%
\providecommand \EOS [0]{\spacefactor3000\relax}%
\providecommand \BibitemShut  [1]{\csname bibitem#1\endcsname}%
\let\auto@bib@innerbib\@empty
\bibitem [{\citenamefont {Abdurrahman}\ \emph {et~al.}(2024)\citenamefont {Abdurrahman}, \citenamefont {Kafker}, \citenamefont {Bulgac},\ and\ \citenamefont {Stetcu}}]{Abdurrahman+2024}%
  \BibitemOpen
  \bibfield  {author} {\bibinfo {author} {\bibfnamefont {I.}~\bibnamefont {Abdurrahman}}, \bibinfo {author} {\bibfnamefont {M.}~\bibnamefont {Kafker}}, \bibinfo {author} {\bibfnamefont {A.}~\bibnamefont {Bulgac}},\ and\ \bibinfo {author} {\bibfnamefont {I.}~\bibnamefont {Stetcu}},\ }\bibfield  {title} {\bibinfo {title} {Neck rupture and scission neutrons in nuclear fission},\ }\href {https://doi.org/10.1103/PhysRevLett.132.242501} {\bibfield  {journal} {\bibinfo  {journal} {Phys. Rev. Lett.}\ }\textbf {\bibinfo {volume} {132}},\ \bibinfo {pages} {242501} (\bibinfo {year} {2024})}\BibitemShut {NoStop}%
\bibitem [{\citenamefont {Andriamirado}\ \emph {et~al.}(2025)\citenamefont {Andriamirado}, \citenamefont {Balantekin}, \citenamefont {Bass}, \citenamefont {Benevides~Rodrigues}, \citenamefont {Bernard}, \citenamefont {Bowden}, \citenamefont {Bryan}, \citenamefont {Carr}, \citenamefont {Classen}, \citenamefont {Conant}, \citenamefont {Deichert}, \citenamefont {Dolinski}, \citenamefont {Erickson}, \citenamefont {Galindo-Uribarri}, \citenamefont {Gokhale}, \citenamefont {Grant}, \citenamefont {Hans}, \citenamefont {Hansell}, \citenamefont {Heeger}, \citenamefont {Heffron}, \citenamefont {Jaffe}, \citenamefont {Jayakumar}, \citenamefont {Koblanski}, \citenamefont {Kunkle}, \citenamefont {Lane}, \citenamefont {Littlejohn}, \citenamefont {Lozano~Sanchez}, \citenamefont {Lu}, \citenamefont {Machado}, \citenamefont {Maricic}, \citenamefont {Mendenhall}, \citenamefont {Meyer}, \citenamefont {Milincic}, \citenamefont {Mueller}, \citenamefont {Mumm}, \citenamefont {Neilson}, \citenamefont {Qian}, \citenamefont {Roca},
  \citenamefont {Rosero}, \citenamefont {Surukuchi}, \citenamefont {Sutanto}, \citenamefont {Venegas-Vargas}, \citenamefont {Weatherly}, \citenamefont {Wilhelmi}, \citenamefont {Yeh}, \citenamefont {Zhang},\ and\ \citenamefont {Zhang}}]{Andriamirado+2025}%
  \BibitemOpen
  \bibfield  {author} {\bibinfo {author} {\bibfnamefont {M.}~\bibnamefont {Andriamirado}}, \bibinfo {author} {\bibfnamefont {A.~B.}\ \bibnamefont {Balantekin}}, \bibinfo {author} {\bibfnamefont {C.~D.}\ \bibnamefont {Bass}}, \bibinfo {author} {\bibfnamefont {O.}~\bibnamefont {Benevides~Rodrigues}}, \bibinfo {author} {\bibfnamefont {E.~P.}\ \bibnamefont {Bernard}}, \bibinfo {author} {\bibfnamefont {N.~S.}\ \bibnamefont {Bowden}}, \bibinfo {author} {\bibfnamefont {C.~D.}\ \bibnamefont {Bryan}}, \bibinfo {author} {\bibfnamefont {R.}~\bibnamefont {Carr}}, \bibinfo {author} {\bibfnamefont {T.}~\bibnamefont {Classen}}, \bibinfo {author} {\bibfnamefont {A.~J.}\ \bibnamefont {Conant}}, \bibinfo {author} {\bibfnamefont {G.}~\bibnamefont {Deichert}}, \bibinfo {author} {\bibfnamefont {M.~J.}\ \bibnamefont {Dolinski}}, \bibinfo {author} {\bibfnamefont {A.}~\bibnamefont {Erickson}}, \bibinfo {author} {\bibfnamefont {A.}~\bibnamefont {Galindo-Uribarri}}, \bibinfo {author} {\bibfnamefont {S.}~\bibnamefont {Gokhale}},
  \bibinfo {author} {\bibfnamefont {C.}~\bibnamefont {Grant}}, \bibinfo {author} {\bibfnamefont {S.}~\bibnamefont {Hans}}, \bibinfo {author} {\bibfnamefont {A.~B.}\ \bibnamefont {Hansell}}, \bibinfo {author} {\bibfnamefont {K.~M.}\ \bibnamefont {Heeger}}, \bibinfo {author} {\bibfnamefont {B.}~\bibnamefont {Heffron}}, \bibinfo {author} {\bibfnamefont {D.~E.}\ \bibnamefont {Jaffe}}, \bibinfo {author} {\bibfnamefont {S.}~\bibnamefont {Jayakumar}}, \bibinfo {author} {\bibfnamefont {J.}~\bibnamefont {Koblanski}}, \bibinfo {author} {\bibfnamefont {P.}~\bibnamefont {Kunkle}}, \bibinfo {author} {\bibfnamefont {C.~E.}\ \bibnamefont {Lane}}, \bibinfo {author} {\bibfnamefont {B.~R.}\ \bibnamefont {Littlejohn}}, \bibinfo {author} {\bibfnamefont {A.}~\bibnamefont {Lozano~Sanchez}}, \bibinfo {author} {\bibfnamefont {X.}~\bibnamefont {Lu}}, \bibinfo {author} {\bibfnamefont {F.}~\bibnamefont {Machado}}, \bibinfo {author} {\bibfnamefont {J.}~\bibnamefont {Maricic}}, \bibinfo {author} {\bibfnamefont {M.~P.}\ \bibnamefont
  {Mendenhall}}, \bibinfo {author} {\bibfnamefont {A.~M.}\ \bibnamefont {Meyer}}, \bibinfo {author} {\bibfnamefont {R.}~\bibnamefont {Milincic}}, \bibinfo {author} {\bibfnamefont {P.~E.}\ \bibnamefont {Mueller}}, \bibinfo {author} {\bibfnamefont {H.~P.}\ \bibnamefont {Mumm}}, \bibinfo {author} {\bibfnamefont {R.}~\bibnamefont {Neilson}}, \bibinfo {author} {\bibfnamefont {X.}~\bibnamefont {Qian}}, \bibinfo {author} {\bibfnamefont {C.}~\bibnamefont {Roca}}, \bibinfo {author} {\bibfnamefont {R.}~\bibnamefont {Rosero}}, \bibinfo {author} {\bibfnamefont {P.~T.}\ \bibnamefont {Surukuchi}}, \bibinfo {author} {\bibfnamefont {F.}~\bibnamefont {Sutanto}}, \bibinfo {author} {\bibfnamefont {D.}~\bibnamefont {Venegas-Vargas}}, \bibinfo {author} {\bibfnamefont {P.~B.}\ \bibnamefont {Weatherly}}, \bibinfo {author} {\bibfnamefont {J.}~\bibnamefont {Wilhelmi}}, \bibinfo {author} {\bibfnamefont {M.}~\bibnamefont {Yeh}}, \bibinfo {author} {\bibfnamefont {C.}~\bibnamefont {Zhang}},\ and\ \bibinfo {author} {\bibfnamefont
  {X.}~\bibnamefont {Zhang}} (\bibinfo {collaboration} {PROSPECT Collaboration}),\ }\bibfield  {title} {\bibinfo {title} {Final search for short-baseline neutrino oscillations with the prospect-i detector at hfir},\ }\href {https://doi.org/10.1103/PhysRevLett.134.151802} {\bibfield  {journal} {\bibinfo  {journal} {Phys. Rev. Lett.}\ }\textbf {\bibinfo {volume} {134}},\ \bibinfo {pages} {151802} (\bibinfo {year} {2025})}\BibitemShut {NoStop}%
\bibitem [{\citenamefont {Wanajo}\ \emph {et~al.}(2024)\citenamefont {Wanajo}, \citenamefont {Fujibayashi}, \citenamefont {Hayashi}, \citenamefont {Kiuchi}, \citenamefont {Sekiguchi},\ and\ \citenamefont {Shibata}}]{Wanajo+2024}%
  \BibitemOpen
  \bibfield  {author} {\bibinfo {author} {\bibfnamefont {S.}~\bibnamefont {Wanajo}}, \bibinfo {author} {\bibfnamefont {S.}~\bibnamefont {Fujibayashi}}, \bibinfo {author} {\bibfnamefont {K.}~\bibnamefont {Hayashi}}, \bibinfo {author} {\bibfnamefont {K.}~\bibnamefont {Kiuchi}}, \bibinfo {author} {\bibfnamefont {Y.}~\bibnamefont {Sekiguchi}},\ and\ \bibinfo {author} {\bibfnamefont {M.}~\bibnamefont {Shibata}},\ }\bibfield  {title} {\bibinfo {title} {Actinide-boosting $r$ process in black-hole--neutron-star merger ejecta},\ }\href {https://doi.org/10.1103/PhysRevLett.133.241201} {\bibfield  {journal} {\bibinfo  {journal} {Phys. Rev. Lett.}\ }\textbf {\bibinfo {volume} {133}},\ \bibinfo {pages} {241201} (\bibinfo {year} {2024})}\BibitemShut {NoStop}%
\bibitem [{\citenamefont {Chadwick}\ \emph {et~al.}(2011)\citenamefont {Chadwick}, \citenamefont {Herman}, \citenamefont {Obložinský}, \citenamefont {Dunn}, \citenamefont {Danon}, \citenamefont {Kahler}, \citenamefont {Smith}, \citenamefont {Pritychenko}, \citenamefont {Arbanas}, \citenamefont {Arcilla}, \citenamefont {Brewer}, \citenamefont {Brown}, \citenamefont {Capote}, \citenamefont {Carlson}, \citenamefont {Cho}, \citenamefont {Derrien}, \citenamefont {Guber}, \citenamefont {Hale}, \citenamefont {Hoblit}, \citenamefont {Holloway}, \citenamefont {Johnson}, \citenamefont {Kawano}, \citenamefont {Kiedrowski}, \citenamefont {Kim}, \citenamefont {Kunieda}, \citenamefont {Larson}, \citenamefont {Leal}, \citenamefont {Lestone}, \citenamefont {Little}, \citenamefont {McCutchan}, \citenamefont {MacFarlane}, \citenamefont {MacInnes}, \citenamefont {Mattoon}, \citenamefont {McKnight}, \citenamefont {Mughabghab}, \citenamefont {Nobre}, \citenamefont {Palmiotti}, \citenamefont {Palumbo}, \citenamefont {Pigni},
  \citenamefont {Pronyaev}, \citenamefont {Sayer}, \citenamefont {Sonzogni}, \citenamefont {Summers}, \citenamefont {Talou}, \citenamefont {Thompson}, \citenamefont {Trkov}, \citenamefont {Vogt}, \citenamefont {van~der Marck}, \citenamefont {Wallner}, \citenamefont {White}, \citenamefont {Wiarda},\ and\ \citenamefont {Young}}]{ENDF}%
  \BibitemOpen
  \bibfield  {author} {\bibinfo {author} {\bibfnamefont {M.~B.}\ \bibnamefont {Chadwick}}, \bibinfo {author} {\bibfnamefont {M.}~\bibnamefont {Herman}}, \bibinfo {author} {\bibfnamefont {P.}~\bibnamefont {Obložinský}}, \bibinfo {author} {\bibfnamefont {M.~E.}\ \bibnamefont {Dunn}}, \bibinfo {author} {\bibfnamefont {Y.}~\bibnamefont {Danon}}, \bibinfo {author} {\bibfnamefont {A.~C.}\ \bibnamefont {Kahler}}, \bibinfo {author} {\bibfnamefont {D.~L.}\ \bibnamefont {Smith}}, \bibinfo {author} {\bibfnamefont {B.}~\bibnamefont {Pritychenko}}, \bibinfo {author} {\bibfnamefont {G.}~\bibnamefont {Arbanas}}, \bibinfo {author} {\bibfnamefont {R.}~\bibnamefont {Arcilla}}, \bibinfo {author} {\bibfnamefont {R.}~\bibnamefont {Brewer}}, \bibinfo {author} {\bibfnamefont {D.~A.}\ \bibnamefont {Brown}}, \bibinfo {author} {\bibfnamefont {R.}~\bibnamefont {Capote}}, \bibinfo {author} {\bibfnamefont {A.~D.}\ \bibnamefont {Carlson}}, \bibinfo {author} {\bibfnamefont {Y.~S.}\ \bibnamefont {Cho}}, \bibinfo {author} {\bibfnamefont
  {H.}~\bibnamefont {Derrien}}, \bibinfo {author} {\bibfnamefont {K.}~\bibnamefont {Guber}}, \bibinfo {author} {\bibfnamefont {G.~M.}\ \bibnamefont {Hale}}, \bibinfo {author} {\bibfnamefont {S.}~\bibnamefont {Hoblit}}, \bibinfo {author} {\bibfnamefont {S.}~\bibnamefont {Holloway}}, \bibinfo {author} {\bibfnamefont {T.~D.}\ \bibnamefont {Johnson}}, \bibinfo {author} {\bibfnamefont {T.}~\bibnamefont {Kawano}}, \bibinfo {author} {\bibfnamefont {B.~C.}\ \bibnamefont {Kiedrowski}}, \bibinfo {author} {\bibfnamefont {H.}~\bibnamefont {Kim}}, \bibinfo {author} {\bibfnamefont {S.}~\bibnamefont {Kunieda}}, \bibinfo {author} {\bibfnamefont {N.~M.}\ \bibnamefont {Larson}}, \bibinfo {author} {\bibfnamefont {L.}~\bibnamefont {Leal}}, \bibinfo {author} {\bibfnamefont {J.~P.}\ \bibnamefont {Lestone}}, \bibinfo {author} {\bibfnamefont {R.~C.}\ \bibnamefont {Little}}, \bibinfo {author} {\bibfnamefont {E.~A.}\ \bibnamefont {McCutchan}}, \bibinfo {author} {\bibfnamefont {R.~E.}\ \bibnamefont {MacFarlane}}, \bibinfo {author}
  {\bibfnamefont {M.}~\bibnamefont {MacInnes}}, \bibinfo {author} {\bibfnamefont {C.~M.}\ \bibnamefont {Mattoon}}, \bibinfo {author} {\bibfnamefont {R.~D.}\ \bibnamefont {McKnight}}, \bibinfo {author} {\bibfnamefont {S.~F.}\ \bibnamefont {Mughabghab}}, \bibinfo {author} {\bibfnamefont {G.~P.~A.}\ \bibnamefont {Nobre}}, \bibinfo {author} {\bibfnamefont {G.}~\bibnamefont {Palmiotti}}, \bibinfo {author} {\bibfnamefont {A.}~\bibnamefont {Palumbo}}, \bibinfo {author} {\bibfnamefont {M.~T.}\ \bibnamefont {Pigni}}, \bibinfo {author} {\bibfnamefont {V.~G.}\ \bibnamefont {Pronyaev}}, \bibinfo {author} {\bibfnamefont {R.~O.}\ \bibnamefont {Sayer}}, \bibinfo {author} {\bibfnamefont {A.~A.}\ \bibnamefont {Sonzogni}}, \bibinfo {author} {\bibfnamefont {N.~C.}\ \bibnamefont {Summers}}, \bibinfo {author} {\bibfnamefont {P.}~\bibnamefont {Talou}}, \bibinfo {author} {\bibfnamefont {I.~J.}\ \bibnamefont {Thompson}}, \bibinfo {author} {\bibfnamefont {A.}~\bibnamefont {Trkov}}, \bibinfo {author} {\bibfnamefont {R.~L.}\
  \bibnamefont {Vogt}}, \bibinfo {author} {\bibfnamefont {S.~C.}\ \bibnamefont {van~der Marck}}, \bibinfo {author} {\bibfnamefont {A.}~\bibnamefont {Wallner}}, \bibinfo {author} {\bibfnamefont {M.~C.}\ \bibnamefont {White}}, \bibinfo {author} {\bibfnamefont {D.}~\bibnamefont {Wiarda}},\ and\ \bibinfo {author} {\bibfnamefont {P.~G.}\ \bibnamefont {Young}},\ }\bibfield  {title} {\bibinfo {title} {{ENDF/B-VII.1 Nuclear Data for Science and Technology: Cross Sections, Covariances, Fission Product Yields and Decay Data}},\ }\href {https://doi.org/10.1016/j.nds.2011.11.002} {\bibfield  {journal} {\bibinfo  {journal} {Nucl. Data Sheets}\ }\textbf {\bibinfo {volume} {112}},\ \bibinfo {pages} {2887} (\bibinfo {year} {2011})},\ \bibinfo {note} {special Issue on ENDF/B-VII.1 Library}\BibitemShut {NoStop}%
\bibitem [{\citenamefont {Iwamoto}\ \emph {et~al.}(2023)\citenamefont {Iwamoto}, \citenamefont {Iwamoto}, \citenamefont {Kunieda}, \citenamefont {Minato}, \citenamefont {Nakayama}, \citenamefont {Abe}, \citenamefont {Tsubakihara}, \citenamefont {Okumura}, \citenamefont {Ishizuka}, \citenamefont {Yoshida}, \citenamefont {Chiba}, \citenamefont {Otuka}, \citenamefont {Sublet}, \citenamefont {Iwamoto}, \citenamefont {Yamamoto}, \citenamefont {Nagaya}, \citenamefont {Tada}, \citenamefont {Konno}, \citenamefont {Matsuda}, \citenamefont {Yokoyama}, \citenamefont {Taninaka}, \citenamefont {Oizumi}, \citenamefont {Fukushima}, \citenamefont {Okita}, \citenamefont {Chiba}, \citenamefont {Sato}, \citenamefont {Ohta},\ and\ \citenamefont {Kwon}}]{jendl23}%
  \BibitemOpen
  \bibfield  {author} {\bibinfo {author} {\bibfnamefont {O.}~\bibnamefont {Iwamoto}}, \bibinfo {author} {\bibfnamefont {N.}~\bibnamefont {Iwamoto}}, \bibinfo {author} {\bibfnamefont {S.}~\bibnamefont {Kunieda}}, \bibinfo {author} {\bibfnamefont {F.}~\bibnamefont {Minato}}, \bibinfo {author} {\bibfnamefont {S.}~\bibnamefont {Nakayama}}, \bibinfo {author} {\bibfnamefont {Y.}~\bibnamefont {Abe}}, \bibinfo {author} {\bibfnamefont {K.}~\bibnamefont {Tsubakihara}}, \bibinfo {author} {\bibfnamefont {S.}~\bibnamefont {Okumura}}, \bibinfo {author} {\bibfnamefont {C.}~\bibnamefont {Ishizuka}}, \bibinfo {author} {\bibfnamefont {T.}~\bibnamefont {Yoshida}}, \bibinfo {author} {\bibfnamefont {S.}~\bibnamefont {Chiba}}, \bibinfo {author} {\bibfnamefont {N.}~\bibnamefont {Otuka}}, \bibinfo {author} {\bibfnamefont {J.-C.}\ \bibnamefont {Sublet}}, \bibinfo {author} {\bibfnamefont {H.}~\bibnamefont {Iwamoto}}, \bibinfo {author} {\bibfnamefont {K.}~\bibnamefont {Yamamoto}}, \bibinfo {author} {\bibfnamefont {Y.}~\bibnamefont
  {Nagaya}}, \bibinfo {author} {\bibfnamefont {K.}~\bibnamefont {Tada}}, \bibinfo {author} {\bibfnamefont {C.}~\bibnamefont {Konno}}, \bibinfo {author} {\bibfnamefont {N.}~\bibnamefont {Matsuda}}, \bibinfo {author} {\bibfnamefont {K.}~\bibnamefont {Yokoyama}}, \bibinfo {author} {\bibfnamefont {H.}~\bibnamefont {Taninaka}}, \bibinfo {author} {\bibfnamefont {A.}~\bibnamefont {Oizumi}}, \bibinfo {author} {\bibfnamefont {M.}~\bibnamefont {Fukushima}}, \bibinfo {author} {\bibfnamefont {S.}~\bibnamefont {Okita}}, \bibinfo {author} {\bibfnamefont {G.}~\bibnamefont {Chiba}}, \bibinfo {author} {\bibfnamefont {S.}~\bibnamefont {Sato}}, \bibinfo {author} {\bibfnamefont {M.}~\bibnamefont {Ohta}},\ and\ \bibinfo {author} {\bibfnamefont {S.}~\bibnamefont {Kwon}},\ }\bibfield  {title} {\bibinfo {title} {{Japanese evaluated nuclear data library version 5: JENDL-5}},\ }\href {https://doi.org/10.1080/00223131.2022.2141903} {\bibfield  {journal} {\bibinfo  {journal} {J. Nucl. Sci. Technol.}\ }\textbf {\bibinfo {volume} {60}},\
  \bibinfo {pages} {1} (\bibinfo {year} {2023})}\BibitemShut {NoStop}%
\bibitem [{\citenamefont {James}\ \emph {et~al.}(1991)\citenamefont {James}, \citenamefont {Mills},\ and\ \citenamefont {Weaver}}]{JEFF}%
  \BibitemOpen
  \bibfield  {author} {\bibinfo {author} {\bibfnamefont {M.}~\bibnamefont {James}}, \bibinfo {author} {\bibfnamefont {R.}~\bibnamefont {Mills}},\ and\ \bibinfo {author} {\bibfnamefont {D.}~\bibnamefont {Weaver}},\ }\bibfield  {title} {\bibinfo {title} {{A new evaluation of fission product yields and the production of a new library (UKFY2) of independent and cumulative yields}},\ }\href {https://doi.org/10.1016/0149-1970(91)90030-S} {\bibfield  {journal} {\bibinfo  {journal} {Prog.\ Nucl.\ Energy}\ }\textbf {\bibinfo {volume} {26}},\ \bibinfo {pages} {1} (\bibinfo {year} {1991})}\BibitemShut {NoStop}%
\bibitem [{\citenamefont {Chen}\ \emph {et~al.}(2024)\citenamefont {Chen}, \citenamefont {Mukobara}, \citenamefont {Ishizuka}, \citenamefont {Katabuchi},\ and\ \citenamefont {Chiba}}]{Chen01}%
  \BibitemOpen
  \bibfield  {author} {\bibinfo {author} {\bibfnamefont {J.}~\bibnamefont {Chen}}, \bibinfo {author} {\bibfnamefont {Y.}~\bibnamefont {Mukobara}}, \bibinfo {author} {\bibfnamefont {C.}~\bibnamefont {Ishizuka}}, \bibinfo {author} {\bibfnamefont {T.}~\bibnamefont {Katabuchi}},\ and\ \bibinfo {author} {\bibfnamefont {S.}~\bibnamefont {Chiba}},\ }\bibfield  {title} {\bibinfo {title} {{Bayesian approach to energy dependence of fission product yields of $^{235}$U by data augmentation}},\ }\href {https://doi.org/10.1080/00223131.2024.2361659} {\bibfield  {journal} {\bibinfo  {journal} {J. Nucl. Sci. Technol.}\ }\textbf {\bibinfo {volume} {61}},\ \bibinfo {pages} {1509} (\bibinfo {year} {2024})}\BibitemShut {NoStop}%
\bibitem [{\citenamefont {{Lovell, Amy}}\ \emph {et~al.}(2019)\citenamefont {{Lovell, Amy}}, \citenamefont {{Mohan, Arvind}}, \citenamefont {{Talou, Patrick}},\ and\ \citenamefont {{Chertkov, Michael}}}]{global_1}%
  \BibitemOpen
  \bibfield  {author} {\bibinfo {author} {\bibnamefont {{Lovell, Amy}}}, \bibinfo {author} {\bibnamefont {{Mohan, Arvind}}}, \bibinfo {author} {\bibnamefont {{Talou, Patrick}}},\ and\ \bibinfo {author} {\bibnamefont {{Chertkov, Michael}}},\ }\bibfield  {title} {\bibinfo {title} {{Constraining Fission Yields Using Machine Learning}},\ }\href {https://doi.org/10.1051/epjconf/201921104006} {\bibfield  {journal} {\bibinfo  {journal} {EPJ Web Conf.}\ }\textbf {\bibinfo {volume} {211}},\ \bibinfo {pages} {04006} (\bibinfo {year} {2019})}\BibitemShut {NoStop}%
\bibitem [{\citenamefont {Wang}\ \emph {et~al.}(2019)\citenamefont {Wang}, \citenamefont {Pei}, \citenamefont {Liu},\ and\ \citenamefont {Qiang}}]{BNN-FPY}%
  \BibitemOpen
  \bibfield  {author} {\bibinfo {author} {\bibfnamefont {Z.-A.}\ \bibnamefont {Wang}}, \bibinfo {author} {\bibfnamefont {J.}~\bibnamefont {Pei}}, \bibinfo {author} {\bibfnamefont {Y.}~\bibnamefont {Liu}},\ and\ \bibinfo {author} {\bibfnamefont {Y.}~\bibnamefont {Qiang}},\ }\bibfield  {title} {\bibinfo {title} {{Bayesian Evaluation of Incomplete Fission Yields}},\ }\href {https://doi.org/10.1103/PhysRevLett.123.122501} {\bibfield  {journal} {\bibinfo  {journal} {Phys. Rev. Lett.}\ }\textbf {\bibinfo {volume} {123}},\ \bibinfo {pages} {122501} (\bibinfo {year} {2019})}\BibitemShut {NoStop}%
\bibitem [{\citenamefont {Tong}\ \emph {et~al.}(2021)\citenamefont {Tong}, \citenamefont {He},\ and\ \citenamefont {Yan}}]{global_3}%
  \BibitemOpen
  \bibfield  {author} {\bibinfo {author} {\bibfnamefont {L.}~\bibnamefont {Tong}}, \bibinfo {author} {\bibfnamefont {R.}~\bibnamefont {He}},\ and\ \bibinfo {author} {\bibfnamefont {S.}~\bibnamefont {Yan}},\ }\bibfield  {title} {\bibinfo {title} {{Prediction of neutron-induced fission product yields by a straightforward $k$-nearest-neighbor algorithm}},\ }\href {https://doi.org/10.1103/PhysRevC.104.064617} {\bibfield  {journal} {\bibinfo  {journal} {Phys. Rev. C}\ }\textbf {\bibinfo {volume} {104}},\ \bibinfo {pages} {064617} (\bibinfo {year} {2021})}\BibitemShut {NoStop}%
\bibitem [{\citenamefont {Lay}\ \emph {et~al.}(2024)\citenamefont {Lay}, \citenamefont {Flynn}, \citenamefont {Giuliani}, \citenamefont {Nazarewicz},\ and\ \citenamefont {Neufcourt}}]{global_4}%
  \BibitemOpen
  \bibfield  {author} {\bibinfo {author} {\bibfnamefont {D.}~\bibnamefont {Lay}}, \bibinfo {author} {\bibfnamefont {E.}~\bibnamefont {Flynn}}, \bibinfo {author} {\bibfnamefont {S.~A.}\ \bibnamefont {Giuliani}}, \bibinfo {author} {\bibfnamefont {W.}~\bibnamefont {Nazarewicz}},\ and\ \bibinfo {author} {\bibfnamefont {L.}~\bibnamefont {Neufcourt}},\ }\bibfield  {title} {\bibinfo {title} {{Neural network emulation of spontaneous fission}},\ }\href {https://doi.org/10.1103/PhysRevC.109.044305} {\bibfield  {journal} {\bibinfo  {journal} {Phys. Rev. C}\ }\textbf {\bibinfo {volume} {109}},\ \bibinfo {pages} {044305} (\bibinfo {year} {2024})}\BibitemShut {NoStop}%
\bibitem [{\citenamefont {Reisdorf}\ \emph {et~al.}(1973)\citenamefont {Reisdorf}, \citenamefont {Unik.},\ and\ \citenamefont {Glendenin}}]{fine_structure}%
  \BibitemOpen
  \bibfield  {author} {\bibinfo {author} {\bibfnamefont {W.}~\bibnamefont {Reisdorf}}, \bibinfo {author} {\bibfnamefont {J.}~\bibnamefont {Unik.}},\ and\ \bibinfo {author} {\bibfnamefont {L.}~\bibnamefont {Glendenin}},\ }\bibfield  {title} {\bibinfo {title} {{Correlation between fragment mass-distribution fine structure, charge distribution and nuclear structure for thermal-neutron-induced fission of $^{233}$U and $^{235}$U}},\ }\href {https://doi.org/https://doi.org/10.1016/0375-9474(73)90214-5} {\bibfield  {journal} {\bibinfo  {journal} {Nucl. Phys. A}\ }\textbf {\bibinfo {volume} {205}},\ \bibinfo {pages} {348} (\bibinfo {year} {1973})}\BibitemShut {NoStop}%
\bibitem [{\citenamefont {Pierson}\ \emph {et~al.}(2017)\citenamefont {Pierson}, \citenamefont {Greenwood}, \citenamefont {Flaska},\ and\ \citenamefont {Pozzi}}]{Pierson_235U_238U_232Th}%
  \BibitemOpen
  \bibfield  {author} {\bibinfo {author} {\bibfnamefont {B.}~\bibnamefont {Pierson}}, \bibinfo {author} {\bibfnamefont {L.}~\bibnamefont {Greenwood}}, \bibinfo {author} {\bibfnamefont {M.}~\bibnamefont {Flaska}},\ and\ \bibinfo {author} {\bibfnamefont {S.}~\bibnamefont {Pozzi}},\ }\bibfield  {title} {\bibinfo {title} {{Fission Product Yields from $^{232}$Th, $^{238}$U, and $^{235}$U Using 14 MeV Neutrons}},\ }\href {https://doi.org/https://doi.org/10.1016/j.nds.2017.01.004} {\bibfield  {journal} {\bibinfo  {journal} {Nucl. Data Sheets}\ }\textbf {\bibinfo {volume} {139}},\ \bibinfo {pages} {171} (\bibinfo {year} {2017})},\ \bibinfo {note} {special Issue on Nuclear Reaction Data}\BibitemShut {NoStop}%
\bibitem [{\citenamefont {Naik}\ \emph {et~al.}(2015)\citenamefont {Naik}, \citenamefont {Mukerji}, \citenamefont {Crasta}, \citenamefont {Suryanarayana}, \citenamefont {Sharma},\ and\ \citenamefont {Goswami}}]{NAIK2015_238U}%
  \BibitemOpen
  \bibfield  {author} {\bibinfo {author} {\bibfnamefont {H.}~\bibnamefont {Naik}}, \bibinfo {author} {\bibfnamefont {S.}~\bibnamefont {Mukerji}}, \bibinfo {author} {\bibfnamefont {R.}~\bibnamefont {Crasta}}, \bibinfo {author} {\bibfnamefont {S.}~\bibnamefont {Suryanarayana}}, \bibinfo {author} {\bibfnamefont {S.}~\bibnamefont {Sharma}},\ and\ \bibinfo {author} {\bibfnamefont {A.}~\bibnamefont {Goswami}},\ }\bibfield  {title} {\bibinfo {title} {{Measurement of fission product yields in the quasi-mono-energetic neutron-induced fission of $^{238}$U}},\ }\href {https://doi.org/https://doi.org/10.1016/j.nuclphysa.2015.05.006} {\bibfield  {journal} {\bibinfo  {journal} {Nucl. Phys. A}\ }\textbf {\bibinfo {volume} {941}},\ \bibinfo {pages} {16} (\bibinfo {year} {2015})}\BibitemShut {NoStop}%
\bibitem [{\citenamefont {Naik}\ \emph {et~al.}(2013)\citenamefont {Naik}, \citenamefont {Mulik}, \citenamefont {Prajapati}, \citenamefont {Shivasankar}, \citenamefont {Suryanarayana}, \citenamefont {Jagadeesan}, \citenamefont {Thakare}, \citenamefont {Sharma},\ and\ \citenamefont {Goswami}}]{NAIK2013_238U}%
  \BibitemOpen
  \bibfield  {author} {\bibinfo {author} {\bibfnamefont {H.}~\bibnamefont {Naik}}, \bibinfo {author} {\bibfnamefont {V.}~\bibnamefont {Mulik}}, \bibinfo {author} {\bibfnamefont {P.}~\bibnamefont {Prajapati}}, \bibinfo {author} {\bibfnamefont {B.}~\bibnamefont {Shivasankar}}, \bibinfo {author} {\bibfnamefont {S.}~\bibnamefont {Suryanarayana}}, \bibinfo {author} {\bibfnamefont {K.}~\bibnamefont {Jagadeesan}}, \bibinfo {author} {\bibfnamefont {S.}~\bibnamefont {Thakare}}, \bibinfo {author} {\bibfnamefont {S.}~\bibnamefont {Sharma}},\ and\ \bibinfo {author} {\bibfnamefont {A.}~\bibnamefont {Goswami}},\ }\bibfield  {title} {\bibinfo {title} {{Mass distribution in the quasi-mono-energetic neutron-induced fission of $^{238}$U}},\ }\href {https://doi.org/https://doi.org/10.1016/j.nuclphysa.2013.05.017} {\bibfield  {journal} {\bibinfo  {journal} {Nucl. Phys. A}\ }\textbf {\bibinfo {volume} {913}},\ \bibinfo {pages} {185} (\bibinfo {year} {2013})}\BibitemShut {NoStop}%
\bibitem [{\citenamefont {Naik}\ \emph {et~al.}(2016)\citenamefont {Naik}, \citenamefont {Mukherji}, \citenamefont {Suryanarayana}, \citenamefont {Jagadeesan}, \citenamefont {Thakare},\ and\ \citenamefont {Sharma}}]{NAIK_2016_232Th}%
  \BibitemOpen
  \bibfield  {author} {\bibinfo {author} {\bibfnamefont {H.}~\bibnamefont {Naik}}, \bibinfo {author} {\bibfnamefont {S.}~\bibnamefont {Mukherji}}, \bibinfo {author} {\bibfnamefont {S.}~\bibnamefont {Suryanarayana}}, \bibinfo {author} {\bibfnamefont {K.}~\bibnamefont {Jagadeesan}}, \bibinfo {author} {\bibfnamefont {S.}~\bibnamefont {Thakare}},\ and\ \bibinfo {author} {\bibfnamefont {S.}~\bibnamefont {Sharma}},\ }\bibfield  {title} {\bibinfo {title} {{Measurement of fission products yields in the quasi-mono-energetic neutron-induced fission of $^{232}$Th}},\ }\href {https://doi.org/https://doi.org/10.1016/j.nuclphysa.2016.04.003} {\bibfield  {journal} {\bibinfo  {journal} {Nucl. Phys. A}\ }\textbf {\bibinfo {volume} {952}},\ \bibinfo {pages} {100} (\bibinfo {year} {2016})}\BibitemShut {NoStop}%
\bibitem [{\citenamefont {Bhatia}\ \emph {et~al.}(2015)\citenamefont {Bhatia}, \citenamefont {Fallin}, \citenamefont {Gooden}, \citenamefont {Howell}, \citenamefont {Kelley}, \citenamefont {Tornow}, \citenamefont {Arnold}, \citenamefont {Bond}, \citenamefont {Bredeweg}, \citenamefont {Fowler}, \citenamefont {Moody}, \citenamefont {Rundberg}, \citenamefont {Rusev}, \citenamefont {Vieira}, \citenamefont {Wilhelmy}, \citenamefont {Becker}, \citenamefont {Macri}, \citenamefont {Ryan}, \citenamefont {Sheets}, \citenamefont {Stoyer},\ and\ \citenamefont {Tonchev}}]{Bhatia_235U_238U_239Pu}%
  \BibitemOpen
  \bibfield  {author} {\bibinfo {author} {\bibfnamefont {C.}~\bibnamefont {Bhatia}}, \bibinfo {author} {\bibfnamefont {B.~F.}\ \bibnamefont {Fallin}}, \bibinfo {author} {\bibfnamefont {M.~E.}\ \bibnamefont {Gooden}}, \bibinfo {author} {\bibfnamefont {C.~R.}\ \bibnamefont {Howell}}, \bibinfo {author} {\bibfnamefont {J.~H.}\ \bibnamefont {Kelley}}, \bibinfo {author} {\bibfnamefont {W.}~\bibnamefont {Tornow}}, \bibinfo {author} {\bibfnamefont {C.~W.}\ \bibnamefont {Arnold}}, \bibinfo {author} {\bibfnamefont {E.}~\bibnamefont {Bond}}, \bibinfo {author} {\bibfnamefont {T.~A.}\ \bibnamefont {Bredeweg}}, \bibinfo {author} {\bibfnamefont {M.~M.}\ \bibnamefont {Fowler}}, \bibinfo {author} {\bibfnamefont {W.}~\bibnamefont {Moody}}, \bibinfo {author} {\bibfnamefont {R.~S.}\ \bibnamefont {Rundberg}}, \bibinfo {author} {\bibfnamefont {G.~Y.}\ \bibnamefont {Rusev}}, \bibinfo {author} {\bibfnamefont {D.~J.}\ \bibnamefont {Vieira}}, \bibinfo {author} {\bibfnamefont {J.~B.}\ \bibnamefont {Wilhelmy}}, \bibinfo {author}
  {\bibfnamefont {J.~A.}\ \bibnamefont {Becker}}, \bibinfo {author} {\bibfnamefont {R.}~\bibnamefont {Macri}}, \bibinfo {author} {\bibfnamefont {C.}~\bibnamefont {Ryan}}, \bibinfo {author} {\bibfnamefont {S.~A.}\ \bibnamefont {Sheets}}, \bibinfo {author} {\bibfnamefont {M.~A.}\ \bibnamefont {Stoyer}},\ and\ \bibinfo {author} {\bibfnamefont {A.~P.}\ \bibnamefont {Tonchev}},\ }\bibfield  {title} {\bibinfo {title} {{Exploratory study of fission product yields of neutron-induced fission of $^{235}\mathrm{U}, ^{238}\mathrm{U}$, and $^{239}\mathrm{Pu}$ at 8.9 MeV}},\ }\href {https://doi.org/10.1103/PhysRevC.91.064604} {\bibfield  {journal} {\bibinfo  {journal} {Phys. Rev. C}\ }\textbf {\bibinfo {volume} {91}},\ \bibinfo {pages} {064604} (\bibinfo {year} {2015})}\BibitemShut {NoStop}%
\bibitem [{\citenamefont {Gooden}\ \emph {et~al.}(2016)\citenamefont {Gooden}, \citenamefont {Arnold}, \citenamefont {Becker}, \citenamefont {Bhatia}, \citenamefont {Bhike}, \citenamefont {Bond}, \citenamefont {Bredeweg}, \citenamefont {Fallin}, \citenamefont {Fowler}, \citenamefont {Howell}, \citenamefont {Kelley}, \citenamefont {Krishichayan}, \citenamefont {Macri}, \citenamefont {Rusev}, \citenamefont {Ryan}, \citenamefont {Sheets}, \citenamefont {Stoyer}, \citenamefont {Tonchev}, \citenamefont {Tornow}, \citenamefont {Vieira},\ and\ \citenamefont {Wilhelmy}}]{Gooden_235U_238U_239Pu}%
  \BibitemOpen
  \bibfield  {author} {\bibinfo {author} {\bibfnamefont {M.}~\bibnamefont {Gooden}}, \bibinfo {author} {\bibfnamefont {C.}~\bibnamefont {Arnold}}, \bibinfo {author} {\bibfnamefont {J.}~\bibnamefont {Becker}}, \bibinfo {author} {\bibfnamefont {C.}~\bibnamefont {Bhatia}}, \bibinfo {author} {\bibfnamefont {M.}~\bibnamefont {Bhike}}, \bibinfo {author} {\bibfnamefont {E.}~\bibnamefont {Bond}}, \bibinfo {author} {\bibfnamefont {T.}~\bibnamefont {Bredeweg}}, \bibinfo {author} {\bibfnamefont {B.}~\bibnamefont {Fallin}}, \bibinfo {author} {\bibfnamefont {M.}~\bibnamefont {Fowler}}, \bibinfo {author} {\bibfnamefont {C.}~\bibnamefont {Howell}}, \bibinfo {author} {\bibfnamefont {J.}~\bibnamefont {Kelley}}, \bibinfo {author} {\bibnamefont {Krishichayan}}, \bibinfo {author} {\bibfnamefont {R.}~\bibnamefont {Macri}}, \bibinfo {author} {\bibfnamefont {G.}~\bibnamefont {Rusev}}, \bibinfo {author} {\bibfnamefont {C.}~\bibnamefont {Ryan}}, \bibinfo {author} {\bibfnamefont {S.}~\bibnamefont {Sheets}}, \bibinfo {author}
  {\bibfnamefont {M.}~\bibnamefont {Stoyer}}, \bibinfo {author} {\bibfnamefont {A.}~\bibnamefont {Tonchev}}, \bibinfo {author} {\bibfnamefont {W.}~\bibnamefont {Tornow}}, \bibinfo {author} {\bibfnamefont {D.}~\bibnamefont {Vieira}},\ and\ \bibinfo {author} {\bibfnamefont {J.}~\bibnamefont {Wilhelmy}},\ }\bibfield  {title} {\bibinfo {title} {{Energy Dependence of Fission Product Yields from $^{235}$U, $^{238}$U and $^{239}$Pu for Incident Neutron Energies Between 0.5 and 14.8 MeV}},\ }\href {https://doi.org/https://doi.org/10.1016/j.nds.2015.12.006} {\bibfield  {journal} {\bibinfo  {journal} {Nucl. Data Sheets}\ }\textbf {\bibinfo {volume} {131}},\ \bibinfo {pages} {319} (\bibinfo {year} {2016})},\ \bibinfo {note} {special Issue on Nuclear Reaction Data}\BibitemShut {NoStop}%
\bibitem [{\citenamefont {Maeck}(1981)}]{burnup-monitor}%
  \BibitemOpen
  \bibfield  {author} {\bibinfo {author} {\bibfnamefont {W.~J.}\ \bibnamefont {Maeck}},\ }\href {https://doi.org/10.2172/5534886} {\emph {\bibinfo {title} {Correlation of $^{239}$Pu thermal and fast reactor fission yields with neutron energy}}}\ (\bibinfo  {publisher} {Exxon Nuclear Idaho Co., Inc., Idaho Falls (USA)},\ \bibinfo {year} {1981})\BibitemShut {NoStop}%
\bibitem [{\citenamefont {Arino}\ and\ \citenamefont {Kramer}(1975)}]{99mTc}%
  \BibitemOpen
  \bibfield  {author} {\bibinfo {author} {\bibfnamefont {H.}~\bibnamefont {Arino}}\ and\ \bibinfo {author} {\bibfnamefont {H.~H.}\ \bibnamefont {Kramer}},\ }\bibfield  {title} {\bibinfo {title} {{Fission product $^{\text{99m}}$Tc generator}},\ }\href {https://doi.org/https://doi.org/10.1016/0020-708X(75)90165-9} {\bibfield  {journal} {\bibinfo  {journal} {Int. J. Appl. Rad. Iso.}\ }\textbf {\bibinfo {volume} {26}},\ \bibinfo {pages} {301} (\bibinfo {year} {1975})}\BibitemShut {NoStop}%
\bibitem [{\citenamefont {Neal}(1996)}]{neal96}%
  \BibitemOpen
  \bibfield  {author} {\bibinfo {author} {\bibfnamefont {R.~M.}\ \bibnamefont {Neal}},\ }\href@noop {} {\emph {\bibinfo {title} {Bayesian Learning for Neural Networks}}}\ (\bibinfo  {publisher} {Springer-Verlag},\ \bibinfo {address} {Berlin, Heidelberg},\ \bibinfo {year} {1996})\BibitemShut {NoStop}%
\bibitem [{\citenamefont {Niu}\ and\ \citenamefont {Liang}(2018)}]{NIU201848}%
  \BibitemOpen
  \bibfield  {author} {\bibinfo {author} {\bibfnamefont {Z.}~\bibnamefont {Niu}}\ and\ \bibinfo {author} {\bibfnamefont {H.}~\bibnamefont {Liang}},\ }\bibfield  {title} {\bibinfo {title} {{Nuclear mass predictions based on Bayesian neural network approach with pairing and shell effects}},\ }\href {https://doi.org/https://doi.org/10.1016/j.physletb.2018.01.002} {\bibfield  {journal} {\bibinfo  {journal} {Phys. Lett. B}\ }\textbf {\bibinfo {volume} {778}},\ \bibinfo {pages} {48} (\bibinfo {year} {2018})}\BibitemShut {NoStop}%
\bibitem [{\citenamefont {M{\"u}ller}\ and\ \citenamefont {f{\"u}r Angewandte~Kernphysik}(1981)}]{mueller1981numerical}%
  \BibitemOpen
  \bibfield  {author} {\bibinfo {author} {\bibfnamefont {R.}~\bibnamefont {M{\"u}ller}}\ and\ \bibinfo {author} {\bibfnamefont {I.}~\bibnamefont {f{\"u}r Angewandte~Kernphysik}},\ }\href {https://books.google.co.jp/books?id=vVaB0AEACAAJ} {\bibinfo {title} {{Numerical Results of a (2E, 2v)-measurement for Fast Neutron Induced Fission of $^{235}$U and $^{237}$Np}}} (\bibinfo {year} {1981})\BibitemShut {NoStop}%
\bibitem [{\citenamefont {Watanabe}(2010{\natexlab{a}})}]{Waic1}%
  \BibitemOpen
  \bibfield  {author} {\bibinfo {author} {\bibfnamefont {S.}~\bibnamefont {Watanabe}},\ }\bibfield  {title} {\bibinfo {title} {{Asymptotic Equivalence of Bayes Cross Validation and Widely Applicable Information Criterion in Singular Learning Theory}},\ }\href {http://jmlr.org/papers/v11/watanabe10a.html} {\bibfield  {journal} {\bibinfo  {journal} {J. Mach. Learn. Res.}\ }\textbf {\bibinfo {volume} {11}},\ \bibinfo {pages} {3571} (\bibinfo {year} {2010}{\natexlab{a}})}\BibitemShut {NoStop}%
\bibitem [{\citenamefont {Hoffman}\ and\ \citenamefont {Gelman}(2011)}]{hmc-nuts}%
  \BibitemOpen
  \bibfield  {author} {\bibinfo {author} {\bibfnamefont {M.~D.}\ \bibnamefont {Hoffman}}\ and\ \bibinfo {author} {\bibfnamefont {A.}~\bibnamefont {Gelman}},\ }\href {https://arxiv.org/abs/1111.4246} {\bibinfo {title} {{The No-U-Turn Sampler: Adaptively Setting Path Lengths in Hamiltonian Monte Carlo}}} (\bibinfo {year} {2011}),\ \Eprint {https://arxiv.org/abs/1111.4246} {arXiv:1111.4246 [stat.CO]} \BibitemShut {NoStop}%
\bibitem [{\citenamefont {Brooks}\ \emph {et~al.}(2011)\citenamefont {Brooks}, \citenamefont {Gelman}, \citenamefont {Jones},\ and\ \citenamefont {Meng}}]{HMC}%
  \BibitemOpen
  \bibfield  {author} {\bibinfo {author} {\bibfnamefont {S.}~\bibnamefont {Brooks}}, \bibinfo {author} {\bibfnamefont {A.}~\bibnamefont {Gelman}}, \bibinfo {author} {\bibfnamefont {G.}~\bibnamefont {Jones}},\ and\ \bibinfo {author} {\bibfnamefont {X.-L.}\ \bibnamefont {Meng}},\ }\href {https://doi.org/10.1201/b10905} {\emph {\bibinfo {title} {{Handbook of Markov Chain Monte Carlo}}}}\ (\bibinfo  {publisher} {Chapman and Hall/CRC},\ \bibinfo {year} {2011})\BibitemShut {NoStop}%
\bibitem [{\citenamefont {Phan}\ \emph {et~al.}(2019)\citenamefont {Phan}, \citenamefont {Pradhan},\ and\ \citenamefont {Jankowiak}}]{numpyro}%
  \BibitemOpen
  \bibfield  {author} {\bibinfo {author} {\bibfnamefont {D.}~\bibnamefont {Phan}}, \bibinfo {author} {\bibfnamefont {N.}~\bibnamefont {Pradhan}},\ and\ \bibinfo {author} {\bibfnamefont {M.}~\bibnamefont {Jankowiak}},\ }\bibfield  {title} {\bibinfo {title} {{Composable Effects for Flexible and Accelerated Probabilistic Programming in NumPyro}},\ }\href@noop {} {\bibfield  {journal} {\bibinfo  {journal} {arXiv preprint arXiv:1912.11554}\ } (\bibinfo {year} {2019})}\BibitemShut {NoStop}%
\bibitem [{\citenamefont {Wilson}\ and\ \citenamefont {Izmailov}(2022)}]{equation5}%
  \BibitemOpen
  \bibfield  {author} {\bibinfo {author} {\bibfnamefont {A.~G.}\ \bibnamefont {Wilson}}\ and\ \bibinfo {author} {\bibfnamefont {P.}~\bibnamefont {Izmailov}},\ }\href {https://arxiv.org/abs/2002.08791} {\bibinfo {title} {Bayesian deep learning and a probabilistic perspective of generalization}} (\bibinfo {year} {2022}),\ \Eprint {https://arxiv.org/abs/2002.08791} {arXiv:2002.08791 [cs.LG]} \BibitemShut {NoStop}%
\bibitem [{\citenamefont {Ishizuka}\ \emph {et~al.}(2017)\citenamefont {Ishizuka}, \citenamefont {Usang}, \citenamefont {Ivanyuk}, \citenamefont {Maruhn}, \citenamefont {Nishio},\ and\ \citenamefont {Chiba}}]{Ishizuka_2017}%
  \BibitemOpen
  \bibfield  {author} {\bibinfo {author} {\bibfnamefont {C.}~\bibnamefont {Ishizuka}}, \bibinfo {author} {\bibfnamefont {M.~D.}\ \bibnamefont {Usang}}, \bibinfo {author} {\bibfnamefont {F.~A.}\ \bibnamefont {Ivanyuk}}, \bibinfo {author} {\bibfnamefont {J.~A.}\ \bibnamefont {Maruhn}}, \bibinfo {author} {\bibfnamefont {K.}~\bibnamefont {Nishio}},\ and\ \bibinfo {author} {\bibfnamefont {S.}~\bibnamefont {Chiba}},\ }\bibfield  {title} {\bibinfo {title} {Four-dimensional langevin approach to low-energy nuclear fission of $^{236}\mathbf{U}$},\ }\href {https://doi.org/10.1103/PhysRevC.96.064616} {\bibfield  {journal} {\bibinfo  {journal} {Phys. Rev. C}\ }\textbf {\bibinfo {volume} {96}},\ \bibinfo {pages} {064616} (\bibinfo {year} {2017})}\BibitemShut {NoStop}%
\bibitem [{\citenamefont {Fujio}\ \emph {et~al.}(2023{\natexlab{a}})\citenamefont {Fujio}, \citenamefont {Al-Adili}, \citenamefont {Nordstr\"om}, \citenamefont {Lema\^itre}, \citenamefont {Okumura}, \citenamefont {Chiba},\ and\ \citenamefont {Koning}}]{fujio2023talys}%
  \BibitemOpen
  \bibfield  {author} {\bibinfo {author} {\bibfnamefont {K.}~\bibnamefont {Fujio}}, \bibinfo {author} {\bibfnamefont {A.}~\bibnamefont {Al-Adili}}, \bibinfo {author} {\bibfnamefont {F.}~\bibnamefont {Nordstr\"om}}, \bibinfo {author} {\bibfnamefont {J.-F.}\ \bibnamefont {Lema\^itre}}, \bibinfo {author} {\bibfnamefont {S.}~\bibnamefont {Okumura}}, \bibinfo {author} {\bibfnamefont {S.}~\bibnamefont {Chiba}},\ and\ \bibinfo {author} {\bibfnamefont {A.}~\bibnamefont {Koning}},\ }\bibfield  {title} {\bibinfo {title} {{TALYS calculations of prompt fission observables and independent fission product yields for the neutron-induced fission of $^{235}$U}},\ }\href {https://doi.org/10.1140/epja/s10050-023-01095-4} {\bibfield  {journal} {\bibinfo  {journal} {Eur. Phys. J. A}\ }\textbf {\bibinfo {volume} {59}},\ \bibinfo {pages} {178} (\bibinfo {year} {2023}{\natexlab{a}})}\BibitemShut {NoStop}%
\bibitem [{\citenamefont {Fujio}\ \emph {et~al.}(2023{\natexlab{b}})\citenamefont {Fujio}, \citenamefont {Okumura}, \citenamefont {Ishizuka}, \citenamefont {Chiba},\ and\ \citenamefont {Katabuchi}}]{fujio2024langevin}%
  \BibitemOpen
  \bibfield  {author} {\bibinfo {author} {\bibfnamefont {K.}~\bibnamefont {Fujio}}, \bibinfo {author} {\bibfnamefont {S.}~\bibnamefont {Okumura}}, \bibinfo {author} {\bibfnamefont {C.}~\bibnamefont {Ishizuka}}, \bibinfo {author} {\bibfnamefont {S.}~\bibnamefont {Chiba}},\ and\ \bibinfo {author} {\bibfnamefont {T.}~\bibnamefont {Katabuchi}},\ }\bibfield  {title} {\bibinfo {title} {{Connection of four-dimensional Langevin model and Hauser-Feshbach theory to describe statistical decay of fission fragments}},\ }\href {https://doi.org/10.1080/00223131.2023.2273470} {\bibfield  {journal} {\bibinfo  {journal} {J. Nucl. Sci. Technol.}\ }\textbf {\bibinfo {volume} {61}},\ \bibinfo {pages} {84} (\bibinfo {year} {2023}{\natexlab{b}})}\BibitemShut {NoStop}%
\bibitem [{\citenamefont {Breiman}\ \emph {et~al.}(1984)\citenamefont {Breiman}, \citenamefont {Friedman}, \citenamefont {Olshen},\ and\ \citenamefont {Stone}}]{k-CV}%
  \BibitemOpen
  \bibfield  {author} {\bibinfo {author} {\bibfnamefont {L.}~\bibnamefont {Breiman}}, \bibinfo {author} {\bibfnamefont {J.}~\bibnamefont {Friedman}}, \bibinfo {author} {\bibfnamefont {R.~A.}\ \bibnamefont {Olshen}},\ and\ \bibinfo {author} {\bibfnamefont {C.~J.}\ \bibnamefont {Stone}},\ }\href {https://doi.org/10.1201/9781315139470} {\emph {\bibinfo {title} {{Classification and Regression Trees}}}},\ \bibinfo {edition} {1st}\ ed.\ (\bibinfo  {publisher} {Chapman and Hall/CRC},\ \bibinfo {year} {1984})\BibitemShut {NoStop}%
\bibitem [{\citenamefont {Watanabe}(2010{\natexlab{b}})}]{watanabe2}%
  \BibitemOpen
  \bibfield  {author} {\bibinfo {author} {\bibfnamefont {S.}~\bibnamefont {Watanabe}},\ }\bibfield  {title} {\bibinfo {title} {{Asymptotic Equivalence of Bayes Cross Validation and Widely Applicable Information Criterion in Singular Learning Theory}},\ }\href {http://jmlr.org/papers/v11/watanabe10a.html} {\bibfield  {journal} {\bibinfo  {journal} {J. Mach. Learn. Res.}\ }\textbf {\bibinfo {volume} {11}},\ \bibinfo {pages} {3571} (\bibinfo {year} {2010}{\natexlab{b}})}\BibitemShut {NoStop}%
\bibitem [{\citenamefont {Chapman}\ \emph {et~al.}(1978)\citenamefont {Chapman}, \citenamefont {Anzelon}, \citenamefont {Spitale},\ and\ \citenamefont {Nethaway}}]{Chapman_235U_238U}%
  \BibitemOpen
  \bibfield  {author} {\bibinfo {author} {\bibfnamefont {T.~C.}\ \bibnamefont {Chapman}}, \bibinfo {author} {\bibfnamefont {G.~A.}\ \bibnamefont {Anzelon}}, \bibinfo {author} {\bibfnamefont {G.~C.}\ \bibnamefont {Spitale}},\ and\ \bibinfo {author} {\bibfnamefont {D.~R.}\ \bibnamefont {Nethaway}},\ }\bibfield  {title} {\bibinfo {title} {{Fission product yeilds from 6-9 MeV neutron-induced fission of $^{235}\mathrm{U}$ and $^{238}\mathrm{U}$}},\ }\href {https://doi.org/10.1103/PhysRevC.17.1089} {\bibfield  {journal} {\bibinfo  {journal} {Phys. Rev. C}\ }\textbf {\bibinfo {volume} {17}},\ \bibinfo {pages} {1089} (\bibinfo {year} {1978})}\BibitemShut {NoStop}%
\bibitem [{\citenamefont {Glendenin}\ \emph {et~al.}(1981)\citenamefont {Glendenin}, \citenamefont {Gindler}, \citenamefont {Henderson},\ and\ \citenamefont {Meadows}}]{Glendenin_235U}%
  \BibitemOpen
  \bibfield  {author} {\bibinfo {author} {\bibfnamefont {L.~E.}\ \bibnamefont {Glendenin}}, \bibinfo {author} {\bibfnamefont {J.~E.}\ \bibnamefont {Gindler}}, \bibinfo {author} {\bibfnamefont {D.~J.}\ \bibnamefont {Henderson}},\ and\ \bibinfo {author} {\bibfnamefont {J.~W.}\ \bibnamefont {Meadows}},\ }\bibfield  {title} {\bibinfo {title} {{Mass distributions for monoenergetic-neutron-induced fission of $^{235}\mathrm{U}$}},\ }\href {https://doi.org/10.1103/PhysRevC.24.2600} {\bibfield  {journal} {\bibinfo  {journal} {Phys. Rev. C}\ }\textbf {\bibinfo {volume} {24}},\ \bibinfo {pages} {2600} (\bibinfo {year} {1981})}\BibitemShut {NoStop}%
\bibitem [{\citenamefont {Selby}\ \emph {et~al.}(2010)\citenamefont {Selby}, \citenamefont {{Mac Innes}}, \citenamefont {Barr}, \citenamefont {Keksis}, \citenamefont {Meade}, \citenamefont {Burns}, \citenamefont {Chadwick},\ and\ \citenamefont {Wallstrom}}]{Selby}%
  \BibitemOpen
  \bibfield  {author} {\bibinfo {author} {\bibfnamefont {H.}~\bibnamefont {Selby}}, \bibinfo {author} {\bibfnamefont {M.}~\bibnamefont {{Mac Innes}}}, \bibinfo {author} {\bibfnamefont {D.}~\bibnamefont {Barr}}, \bibinfo {author} {\bibfnamefont {A.}~\bibnamefont {Keksis}}, \bibinfo {author} {\bibfnamefont {R.}~\bibnamefont {Meade}}, \bibinfo {author} {\bibfnamefont {C.}~\bibnamefont {Burns}}, \bibinfo {author} {\bibfnamefont {M.}~\bibnamefont {Chadwick}},\ and\ \bibinfo {author} {\bibfnamefont {T.}~\bibnamefont {Wallstrom}},\ }\bibfield  {title} {\bibinfo {title} {{Fission Product Data Measured at Los Alamos for Fission Spectrum and Thermal Neutrons on $^{239}$Pu, $^{235}$U, $^{238}$U}},\ }\href {https://doi.org/https://doi.org/10.1016/j.nds.2010.11.002} {\bibfield  {journal} {\bibinfo  {journal} {Nucl. Data Sheets}\ }\textbf {\bibinfo {volume} {111}},\ \bibinfo {pages} {2891} (\bibinfo {year} {2010})},\ \bibinfo {note} {nuclear Reaction Data}\BibitemShut {NoStop}%
\bibitem [{\citenamefont {Laurec}\ \emph {et~al.}(2010)\citenamefont {Laurec}, \citenamefont {Adam}, \citenamefont {{de Bruyne}}, \citenamefont {Bauge}, \citenamefont {Granier}, \citenamefont {Aupiais}, \citenamefont {Bersillon}, \citenamefont {{Le Petit}}, \citenamefont {Authier},\ and\ \citenamefont {Casoli}}]{Laurec_233U_235U_239Pu}%
  \BibitemOpen
  \bibfield  {author} {\bibinfo {author} {\bibfnamefont {J.}~\bibnamefont {Laurec}}, \bibinfo {author} {\bibfnamefont {A.}~\bibnamefont {Adam}}, \bibinfo {author} {\bibfnamefont {T.}~\bibnamefont {{de Bruyne}}}, \bibinfo {author} {\bibfnamefont {E.}~\bibnamefont {Bauge}}, \bibinfo {author} {\bibfnamefont {T.}~\bibnamefont {Granier}}, \bibinfo {author} {\bibfnamefont {J.}~\bibnamefont {Aupiais}}, \bibinfo {author} {\bibfnamefont {O.}~\bibnamefont {Bersillon}}, \bibinfo {author} {\bibfnamefont {G.}~\bibnamefont {{Le Petit}}}, \bibinfo {author} {\bibfnamefont {N.}~\bibnamefont {Authier}},\ and\ \bibinfo {author} {\bibfnamefont {P.}~\bibnamefont {Casoli}},\ }\bibfield  {title} {\bibinfo {title} {{Fission Product Yields of $^{233}$U, $^{235}$U, $^{238}$U and $^{239}$Pu in Fields of Thermal Neutrons, Fission Neutrons and 14.7-MeV Neutrons}},\ }\href {https://doi.org/https://doi.org/10.1016/j.nds.2010.11.004} {\bibfield  {journal} {\bibinfo  {journal} {Nucl. Data Sheets}\ }\textbf {\bibinfo {volume} {111}},\
  \bibinfo {pages} {2965} (\bibinfo {year} {2010})},\ \bibinfo {note} {nuclear Reaction Data}\BibitemShut {NoStop}%
\bibitem [{\citenamefont {Ivanyuk}\ \emph {et~al.}(2025)\citenamefont {Ivanyuk}, \citenamefont {Schmitt}, \citenamefont {Ishizuka},\ and\ \citenamefont {Chiba}}]{multichance_fission}%
  \BibitemOpen
  \bibfield  {author} {\bibinfo {author} {\bibfnamefont {F.~A.}\ \bibnamefont {Ivanyuk}}, \bibinfo {author} {\bibfnamefont {C.}~\bibnamefont {Schmitt}}, \bibinfo {author} {\bibfnamefont {C.}~\bibnamefont {Ishizuka}},\ and\ \bibinfo {author} {\bibfnamefont {S.}~\bibnamefont {Chiba}},\ }\href {https://arxiv.org/abs/2501.05362} {\bibinfo {title} {{Shell effects and multi-chance fission in the sub-lead region}}} (\bibinfo {year} {2025}),\ \Eprint {https://arxiv.org/abs/2501.05362} {arXiv:2501.05362 [nucl-th]} \BibitemShut {NoStop}%
\bibitem [{\citenamefont {Brosa}\ \emph {et~al.}(1990)\citenamefont {Brosa}, \citenamefont {Grossmann},\ and\ \citenamefont {Müller}}]{brosa}%
  \BibitemOpen
  \bibfield  {author} {\bibinfo {author} {\bibfnamefont {U.}~\bibnamefont {Brosa}}, \bibinfo {author} {\bibfnamefont {S.}~\bibnamefont {Grossmann}},\ and\ \bibinfo {author} {\bibfnamefont {A.}~\bibnamefont {Müller}},\ }\bibfield  {title} {\bibinfo {title} {Nuclear scission},\ }\href {https://doi.org/10.1016/0370-1573(90)90114-H} {\bibfield  {journal} {\bibinfo  {journal} {Phys.\ Rep.}\ }\textbf {\bibinfo {volume} {197}},\ \bibinfo {pages} {167} (\bibinfo {year} {1990})}\BibitemShut {NoStop}%
\bibitem [{\citenamefont {Scamps}\ and\ \citenamefont {Simenel}(2018)}]{nature1}%
  \BibitemOpen
  \bibfield  {author} {\bibinfo {author} {\bibfnamefont {G.}~\bibnamefont {Scamps}}\ and\ \bibinfo {author} {\bibfnamefont {C.}~\bibnamefont {Simenel}},\ }\bibfield  {title} {\bibinfo {title} {{Impact of pear-shaped fission fragments on mass-asymmetric fission in actinides}},\ }\href {https://doi.org/10.1038/s41586-018-0780-0} {\bibfield  {journal} {\bibinfo  {journal} {Nature}\ }\textbf {\bibinfo {volume} {564}},\ \bibinfo {pages} {382} (\bibinfo {year} {2018})}\BibitemShut {NoStop}%
\bibitem [{\citenamefont {Huo}\ \emph {et~al.}(2023)\citenamefont {Huo}, \citenamefont {Wei}, \citenamefont {Wu}, \citenamefont {Han}, \citenamefont {Han}, \citenamefont {Wang}, \citenamefont {Zhang}, \citenamefont {He}, \citenamefont {Bao}, \citenamefont {Deng},\ and\ \citenamefont {Yao}}]{explanation_huo}%
  \BibitemOpen
  \bibfield  {author} {\bibinfo {author} {\bibfnamefont {D.}~\bibnamefont {Huo}}, \bibinfo {author} {\bibfnamefont {Z.}~\bibnamefont {Wei}}, \bibinfo {author} {\bibfnamefont {K.}~\bibnamefont {Wu}}, \bibinfo {author} {\bibfnamefont {C.}~\bibnamefont {Han}}, \bibinfo {author} {\bibfnamefont {Y.-n.}\ \bibnamefont {Han}}, \bibinfo {author} {\bibfnamefont {Y.-x.}\ \bibnamefont {Wang}}, \bibinfo {author} {\bibfnamefont {P.-q.}\ \bibnamefont {Zhang}}, \bibinfo {author} {\bibfnamefont {Y.}~\bibnamefont {He}}, \bibinfo {author} {\bibfnamefont {X.-j.}\ \bibnamefont {Bao}}, \bibinfo {author} {\bibfnamefont {Z.-y.}\ \bibnamefont {Deng}},\ and\ \bibinfo {author} {\bibfnamefont {Z.-e.}\ \bibnamefont {Yao}},\ }\bibfield  {title} {\bibinfo {title} {{Effect of octupole deformation of fragments on mass-asymmetric yields of fission of actinide nuclei}},\ }\href {https://doi.org/10.1103/PhysRevC.108.024608} {\bibfield  {journal} {\bibinfo  {journal} {Phys. Rev. C}\ }\textbf {\bibinfo {volume} {108}},\ \bibinfo {pages} {024608}
  (\bibinfo {year} {2023})}\BibitemShut {NoStop}%
\bibitem [{\citenamefont {Tonchev}\ \emph {et~al.}(2025)\citenamefont {Tonchev}, \citenamefont {Silano}, \citenamefont {Ramirez}, \citenamefont {Malone}, \citenamefont {Stoyer}, \citenamefont {Gooden}, \citenamefont {Bredeweg}, \citenamefont {Vieira}, \citenamefont {Wilhelmy}, \citenamefont {Finch}, \citenamefont {Howell},\ and\ \citenamefont {Tornow}}]{Tonchev2025}%
  \BibitemOpen
  \bibfield  {author} {\bibinfo {author} {\bibfnamefont {A.}~\bibnamefont {Tonchev}}, \bibinfo {author} {\bibfnamefont {J.}~\bibnamefont {Silano}}, \bibinfo {author} {\bibfnamefont {A.}~\bibnamefont {Ramirez}}, \bibinfo {author} {\bibfnamefont {R.}~\bibnamefont {Malone}}, \bibinfo {author} {\bibfnamefont {M.}~\bibnamefont {Stoyer}}, \bibinfo {author} {\bibfnamefont {M.}~\bibnamefont {Gooden}}, \bibinfo {author} {\bibfnamefont {T.}~\bibnamefont {Bredeweg}}, \bibinfo {author} {\bibfnamefont {D.}~\bibnamefont {Vieira}}, \bibinfo {author} {\bibfnamefont {J.}~\bibnamefont {Wilhelmy}}, \bibinfo {author} {\bibfnamefont {S.}~\bibnamefont {Finch}}, \bibinfo {author} {\bibfnamefont {C.}~\bibnamefont {Howell}},\ and\ \bibinfo {author} {\bibfnamefont {W.}~\bibnamefont {Tornow}},\ }\bibfield  {title} {\bibinfo {title} {{Energy dependence of chain fission product yields from neutron-induced fission of $^{235}$U, $^{238}$U, and $^{239}$Pu}},\ }\href {https://doi.org/https://doi.org/10.1016/j.nds.2025.04.002} {\bibfield
  {journal} {\bibinfo  {journal} {Nucl. Data Sheets}\ }\textbf {\bibinfo {volume} {202}},\ \bibinfo {pages} {12} (\bibinfo {year} {2025})}\BibitemShut {NoStop}%
\end{thebibliography}%

\end{document}